\documentclass[useAMS,usenatbib]{mnras}
\usepackage{amsmath}

\input psfig.sty

\usepackage{graphicx}
\usepackage{epstopdf}
\usepackage{dsfont}

\renewcommand{\vec}[1]{ {\bf #1} }

\def\spose#1{\hbox to 0pt{#1\hss}}
\def\lta{\mathrel{\spose{\lower 3pt\hbox{$\mathchar"218$}}
     \raise 2.0pt\hbox{$\mathchar"13C$}}}
\def\gta{\mathrel{\spose{\lower 3pt\hbox{$\mathchar"218$}}
     \raise 2.0pt\hbox{$\mathchar"13E$}}}
\def\kms{{km s$^{-1}$}}

\def \chisq  {\ifmmode  \chi^2   \else  $\chi^2$  \fi}  
\def \spose#1{\hbox  to 0pt{#1\hss}}  
\def \lta{\mathrel{\spose{\lower 3pt\hbox{$\sim$}}\raise  2.0pt\hbox{$<$}}}
\def \gta{\mathrel{\spose{\lower  3pt\hbox{$\sim$}}\raise 2.0pt\hbox{$>$}}}

\def \kms {\ifmmode  \,\rm km\,s^{-1} \else $\,\rm km\,s^{-1}  $ \fi }
\def \kpc {\ifmmode  {\rm~kpc}  \else ${\rm~kpc}$\fi}  
\def \pc {\ifmmode  {\rm~pc}  \else ${\rm~pc}$ \fi  }  
\def \Gyr {\ifmmode  {\rm~Gyr}  \else ${\rm~Gyr}$\fi}
\def \Msun {\ifmmode \rm{M}_{\odot} \else M$_{\odot}$ \fi} 
\def \Lsun {\ifmmode L_{\odot} \else $L_{\odot}$ \fi} 
\def \Rsun {\ifmmode R_{\odot} \else $R_{\odot}$ \fi} 
\def \Msunpyr {\ifmmode M_{\odot}{\rm~yr}^{-1} \else $M_{\odot}{\rm~yr}^{-1}$ \fi} 
\def \hMsun {\ifmmode h^{-1}\,\rm M_{\odot} \else $h^{-1}\,\rm M_{\odot}$ \fi}

\def \LCDM {\ifmmode \Lambda{\rm CDM} \else $\Lambda{\rm CDM}$ \fi}
\def \sig8 {\ifmmode \sigma_8 \else $\sigma_8$ \fi} 
\def \OmegaM {\ifmmode \Omega_{\rm M} \else $\Omega_{\rm M}$ \fi} 
\def \OmegaL {\ifmmode \Omega_{\rm \Lambda} \else $\Omega_{\rm \Lambda}$\fi} 
\def \Omegab {\ifmmode \Omega_{\rm b} \else $\Omega_{\rm b}$ \fi}
\def \Deltavir {\ifmmode \Delta_{\rm vir} \else $\Delta_{\rm vir}$ \fi}
\def \rhocrit {\ifmmode \rho_{\rm crit} \else $\rho_{\rm crit}$ \fi}
\def \rhou {\ifmmode \rho_{\rm u} \else $\rho_{\rm u}$ \fi}
\def \zc {\ifmmode z_{\rm c} \else $z_{\rm c}$ \fi}

\def \rhos {\ifmmode \rho_{\rm s} \else $\rho_{\rm s}$ \fi} 
\def \rs {\ifmmode r_{\rm s} \else $r_{\rm s}$ \fi} 
\def \cvir {\ifmmode c_{\rm vir} \else $c_{\rm vir}$ \fi} 
\def \Rvir {\ifmmode r_{\rm vir} \else $R_{\rm vir}$ \fi}
\def \Vvir {\ifmmode V_{\rm  vir} \else  $V_{\rm vir}$  \fi} 
\def \Mvir {\ifmmode M_{\rm  vir} \else $M_{\rm  vir}$ \fi}  
\def \Nvir {\ifmmode N_{\rm  vir} \else $N_{\rm  vir}$ \fi}  
\def \Jvir {\ifmmode J_{\rm vir} \else $J_{\rm vir}$ \fi} 
\def \Evir {\ifmmode E_{\rm vir} \else $E_{\rm vir}$ \fi} 
\def \vvir {\ifmmode v_{\rm vir} \else $v_{\rm vir}$ \fi} 
\def \lam {\ifmmode \lambda  \else $\lambda$ \fi} 
\def \lamp {\ifmmode \lambda^{\prime} \else $\lambda^{\prime}$  \fi} 
\def \Vmax {\ifmmode V_{\rm  max} \else  $V_{\rm max}$  \fi} 
\def \Mdm {\ifmmode M_{\rm  dm} \else $M_{\rm  dm}$ \fi}

\def \Mgas {\ifmmode M_{\rm gas} \else $M_{\rm gas}$ \fi} 
\def \Mcg {\ifmmode M_{\rm cg} \else $M_{\rm cg}$\fi} 
\def \Mhg {\ifmmode M_{\rm hg} \else $M_{\rm hg}$ \fi} 
\def \Mdisc {\ifmmode M_{\rm disc} \else $M_{\rm disc}$ \fi} 
\def \Md {\ifmmode M_{\rm d} \else $M_{\rm d}$ \fi} 
\def \Mda {\ifmmode M_{\rm d,0\%} \else $M_{\rm d,0\%}$ \fi} 
\def \Mdb {\ifmmode M_{\rm d,20\%} \else $M_{\rm d,20\%}$ \fi} 
\def \Mdc {\ifmmode M_{\rm d,40\%} \else $M_{\rm d,40\%}$ \fi} 
\def \md {\ifmmode m_{\rm d} \else $m_{\rm d}$ \fi} 
\def \Mb {\ifmmode M_{\rm b} \else $M_{\rm b}$ \fi} 
\def \Mbh {\ifmmode M_{\rm b,pri} \else $M_{\rm b,pri}$ \fi} 
\def \Mbs {\ifmmode M_{\rm b,sat} \else $M_{\rm b,sat}$ \fi} 
\def \zo {\ifmmode z_{0} \else $z_{0}$ \fi} 

\def \rb {\ifmmode r_{\rm b} \else $r_{\rm b}$\fi}
\def \rs {\ifmmode r_{\rm s} \else $r_{\rm s}$\fi}
\def \rc {\ifmmode r_{\rm c} \else $r_{\rm c}$\fi}
\def \rvir {\ifmmode r_{\rm vir} \else $r_{\rm vir}$\fi}
\def \rbh {\ifmmode r_{\rm b,pri} \else $r_{\rm b,pri}$ \fi} 
\def \rbs {\ifmmode r_{\rm b,sat} \else $r_{\rm b,sat}$ \fi}

\title[Anisotropic thermal conduction]{Accurately simulating anisotropic thermal conduction on a moving mesh}

\author[Kannan et al.]{Rahul Kannan$^1$\thanks{Email: kannanr@mit.edu}, Volker Springel$^{2,3}$, R\"udiger Pakmor$^{2}$, Federico Marinacci$^1$, \newauthor{Mark Vogelsberger$^1$}
\vspace*{6pt}\\
$^1$Department of Physics, Kavli Institute for Astrophysics $\&$ Space Research, Massachusetts Institute of Technology, Cambridge 02139, MA, USA \\
$^2$Heidelberg Institute for Theoretical Studies, Schloss-Wolfsbrunnenweg 35, D-69118 Heidelberg, Germany \\
$^3$Zentrum f\"ur Astronomie der Universit\"at Heidelberg, ARI, M\"onchhofstr. 12-14, D-69120 Heidelberg, Germany}

\begin{document}

\maketitle

\pagerange{\pageref{firstpage}--\pageref{lastpage}}
\pubyear{2015}

\label{firstpage}

\begin{abstract} 
  We present a novel implementation of an extremum preserving anisotropic
  diffusion solver for thermal conduction on the unstructured moving
  Voronoi mesh of the {\sc Arepo} code. The method relies on splitting
  the one-sided facet fluxes into normal and oblique components, with
  the oblique fluxes being limited such that the total flux is both locally
  conservative and extremum preserving. The approach makes use of
  harmonic averaging points and a simple, robust interpolation scheme
  that works well for strong heterogeneous and anisotropic diffusion
  problems. Moreover, the required discretisation stencil is
  small. Efficient fully implicit and semi-implicit time integration
  schemes are also implemented.  We perform several numerical tests
  that evaluate the stability and accuracy of the scheme, including
  applications such as point explosions with heat conduction and
  calculations of convective instabilities in conducting plasmas.  The
  new implementation is suitable for studying important astrophysical
  phenomena, such as the conductive heat transport in galaxy clusters,
  the evolution of supernova remnants, or the distribution of heat
  from blackhole-driven jets into the intracluster medium.
\end{abstract}


\section{Introduction}

Diffusion processes are ubiquitous in Nature. They describe the
transport of a substance or quantity down a gradient, and anisotropy
in this transport arises when the rate of diffusion varies in
particular directions.  Such anisotropic diffusion occurs not only in
astronomy but in fact appears in a wide range of scientific fields,
including biological systems \citep{Biology2014}, image processing
\citep{Weickert1998} plasma physics \citep{Plasma2014}, and 
petroleum reservoir simulations \citep{Petroleum2005}.  In
astrophysics, the anisotropy is mainly caused by the presence of
magnetic fields which force charged particles, such as cosmic rays
(CRs) \citep{Grinzburg1980} and electrons \citep{Spitzer1962}, to move
primarily along the magnetic field lines.

CRs are high-energy relativistic particles ($>10^9 \ {\rm eV}$) that
exhibit a nearly featureless power-law energy spectrum over eleven
orders of magnitude \citep{CR2002}.  They are thought to be generated
by acceleration of charged particles in supernova remnants, active
galactic nuclei (AGN) and gamma-ray bursts \citep{Fermi1949}.  CRs
appear to be a very important part of our galaxy ecosystem, as their
energy density is comparable to the energy density of interstellar
magnetic fields, of diffuse starlight, and of the kinetic energy
density of interstellar gas. This provides an important hint about the
interplay between all these components. In fact, CR have been
suggested to play an important role in regulating star formation in
galaxies and in driving galactic winds and outflows
\citep{Jubelgas2008, Booth2013, Pfrommer2013}. 

Thermal conduction is the process through which internal energy is
diffusively transported by collision of particles. In high-energy
plasmas, electrons are the primary carriers for heat transfer. The thermal conductivity is a strong function of temperature
($\kappa \propto T^{5/2}$) and hence the conduction timescales are
comparable to the dynamical timescales of the system only in a high
temperature plasma \citep{Spitzer1962}, as for instance in galaxy
clusters \citep{Voit2015N}, or supernova winds \citep{  Balsara2008, Balsara2008b, Thompson2015}. Conduction has been invoked by many
authors to explain the low radiative cooling rate in clusters
\citep{Zakamska2003, Voit2011, Voit2015}, in terms of a
conductive heat flow that offsets the central cooling losses.
However, in the presence of magnetic fields, electrons preferentially
move along magnetic field lines, and hence heat transport becomes
anisotropic. This complicates the theoretical modelling considerably.

Another interesting phenomenon is the generation of turbulent pressure
support \citep{McCourt2013} in clusters induced by buoyancy
instabilities coupled to thermal conduction \citep{Balbus2000,
  Parrish2005, Parrish2008, Sharma2008, McCourt2011}. These
instabilities can exponentially amplify magnetic fields and may be
responsible for the observed ${\rm \mu G}$ magnetic fields in galaxy
clusters \citep{Quataert2008}.

Simulating these interesting phenomena of course requires a stable
numerical implementation of an anisotropic diffusion
solver. Unfortunately, the discretisation of the anisotropic diffusion
equation is surprisingly non-trivial. Widely used discretisation
approaches \citep{Parrish2005, Balsara2008, Petkova2009, Arth2014,
  Dubois2015} do not satisfy the so-called discrete extremum principle
(DEP), which is the discrete version of the extremum principle defined
as follows. Let $u_o(\vec{x})$ be the spatial distribution of the
quantity $u$ within the domain at the starting time. We define
\begin{eqnarray}
  M &= &\max(u_0(\vec{x}), \\
 m & =&  \min(u_0(\vec{x})).
\end{eqnarray}
The  discrete extremum principle then states that
\begin{equation}
  u(\vec{x}, t) \in [m, M] \ \forall  \  \vec{x}, t .
\end{equation}
The discrete extremum principle hence ensures that values of the
variable $u(\vec{x},t)$ in the evolving system are constrained to lie
within the range covered by the initial values ($u_0(\vec{x})$). It is
a very restrictive condition that guarantees that over- and
under-shoots cannot appear.

Another closely related concept is monotonicity preservation, which
usually means that the scheme is non-negativity maintaining. For linear
diffusion problems, the DEP and monotonicity preservation are
equivalent \citep{Sheng2011}.  \citet{Pert1981} have pointed out that
schemes violating DEP can create spurious negative values in the
solution which can cause unphysical oscillations. In the context of
thermal conduction, disobeying DEP/monotonicity can lead to a
violation of the the second law of thermodynamics, causing heat to
flow from regions of lower temperature to areas of higher
temperature. In general cases (including the nonlinear cases which we
consider), the discrete extremum principle is more restrictive than
monotonicity.

Many works have tried to formulate stable anisotropic diffusion
solvers. \citet{Petkova2009} and \citet{Arth2014} implemented a
(mildly) isotropised version of an anisotropic diffusion scheme based
on smooth particle hydrodynamics (SPH). Their method relies on adding
an additional isotropic component to the anisotropic diffusion tensor
in order to avoid unphysical oscillations. However, a significant
amount ($\sim 60\%$ of the total diffusion flux) of isotropic
diffusivity needs to be added in order to achieve full numerical
stability. This situation is clearly undesirable and can significantly
affect the results of simulations \citep{Petkova2009}.

There are several linear schemes which do satisfy the DEP, but they
impose severe restrictions on the allowed topology of the meshes
and/or the diffusion coefficients \citep{Aavatsmark1996,
  Breil2007}. For example, \citet{Sharma2007} proposed a class of
slope-limited methods for anisotropic thermal conduction which are
monotonicity preserving. They decomposed the temperature gradient
into two components: the normal term and the transverse term. The
normal term always gives flux from higher to lower temperature, but
the transverse term can be of any sign. Therefore, the transverse term
is slope-limited to ensure that extrema are not
accentuated. \citet{Rasera2008} use the so-called flux tube method for
this purpose. It discretizes anisotropic diffusion as a 1D problem
along the magnetic field line and then re-projects the result on to
the grid in such a way that it is positivity preserving. These methods
work very well for regular cartesian grids.  However, their extension
to unstructured meshes is non-trivial.

Many non-linear methods have been proposed to solve this problem. In a
non-linear scheme, the coefficients of the scheme depend on the
solution itself. The non-linear method proposed in \citep{Potier2005}
satisfies either the discrete minimum or maximum principle but not
both. Further improvements were made to this method by
\citet{Yuan2008, Sheng2011, Sheng2012}.  The resulting class of
methods relies on determining a suitable approximation to the cell face
quantities in addition to the cell centered ones. Therefore, an
interpolation scheme is required in order to interpolate the cell
centered unknowns onto the cell faces.  However, the interpolation
schemes proposed in these works were not proven to be positivity
preserving, which is an important condition in order to follow the
DEP.

In this paper, we follow the method outlined in \citealt{Gao2013}
(GW13 from hereon) to implement an extremum preserving anisotropic
diffusion solver for heat conduction on moving Voronoi meshes. The
scheme makes use of the harmonic averaging points suggested in
\citet{Agelas2009} and \citet{Eymard2012}, and proposes a very simple
yet robust interpolation scheme that works well for strong
heterogeneous and anisotropic diffusion problems. In addition, the
required discretisation stencil is essentially small; in our case is
consists of the cell and all its Delaunay connections (see Section~3
for details). This makes it easy to implement on the unstructured
moving Voronoi mesh used in the {\sc Arepo} code \citep{Springel2010}.
The use of harmonic averaging points also makes the interpolation
positivity preserving. GW13 prove that this method is locally
conservative and follows the DEP under the assumption that the cell is
convex in nature. This condition is always satisfied in {\sc Arepo},
because Voronoi meshes are convex by construction.

The paper is organized as follows. In Section~2, we outline the basic
equations of anisotropic thermal conduction.  Section~3 describes the
spatial discretisation, interpolation techniques and the time
integration methods that we use. In Section~4 we show several test
problems and asses the validity and accuracy of our
algorithm. Finally, we present our conclusions in Section~5.

\section{Thermal Conduction}
\label{tc:theory}

We start with the energy conservation equation which can be written
as
 \begin{equation}
 \rho \frac{\partial u}{\partial t}  + \vec{\nabla} \cdot \vec{j} = 0,
 \label{dfeq}
\end{equation}
where $u$ is the gas internal energy per unit mass, $\vec{j}$ is the 
directional heat flux and $\rho$ is the gas density. If the conduction is 
isotropic, then the heat flux $\vec{j}$ is opposite to the direction of the 
temperature gradient
\begin{equation}
\vec{j} = -\kappa_{\rm sp}(T)\, \vec{\nabla}T.
\end{equation}
 $\kappa_{\rm sp}$ is the conduction coefficient given by \citet{Spitzer1962},
 \begin{equation}
\kappa_{\rm sp} = \frac{1.84 \times 10^{-5} T^{5/2}}{\ln \  C} \ {\rm ergs \ s^{-1} \ K^{-1} \ cm^{-1}},
\label{eq:ksp}
\end{equation}
where $\ln \ C \sim 37$ is the so-called Coulomb logarithm and 
$T$ is the gas temperature. 

This value of the conduction coefficient is only valid under the
assumption that the typical length scale of the temperature gradient,
$l_T = T/|\delta T|$, is much larger than the mean free path $l_e$ of
the electrons. This assumption breaks down at very low densities or
for extreme temperature gradients. In such cases, it is estimated that
the heat flux saturates to a much lower value given by
\citep{Cowie1977}
\begin{equation}
 j_{\rm sat} \sim 0.4 n_e k_B T \left( \frac{2k_{\rm B} T}{\pi m_e} \right)^{\frac{1}{2}},
\end{equation}
where $n_e$ is the number density of electrons, $k_{\rm B}$ is the
Boltzmann constant and $m_e$ is the mass of the electron.

In order to capture this behavior and achieve a smooth transition
between the Spitzer and the saturated regimes, we modify the conduction
coefficient and set it to \citep{Sarazin1988}
\begin{equation}
 \kappa = \frac{\kappa_{\rm sp}}{1 + 4.2l_e/l_T},
\end{equation}
where 
\begin{equation}
l_e = \frac{3^{3/2} (k_B T)^2}{4 n_e \sqrt{\pi}e^4 \ ln \ C}
\end{equation}
and $e$ denotes the electronic charge. 

In the presence of the magnetic fields, electrons preferentially move
along the magnetic field lines. Therefore the heat flux is modified as
\begin{equation}
\vec{j} = -\kappa[\vec{b}(\vec{b} \cdot \vec{\nabla}T)] ,
\label{eq:hfatc}
\end{equation}
where $\vec{b} = \vec{B}/|\vec{B}|$. The coefficient of conduction
perpendicular to the magnetic fields is set to zero (see
\citealt{Arth2014} for a discussion on non vanishing perpendicular
conduction coefficients).

\begin{figure*}
\begin{center}
\includegraphics[scale=0.5]{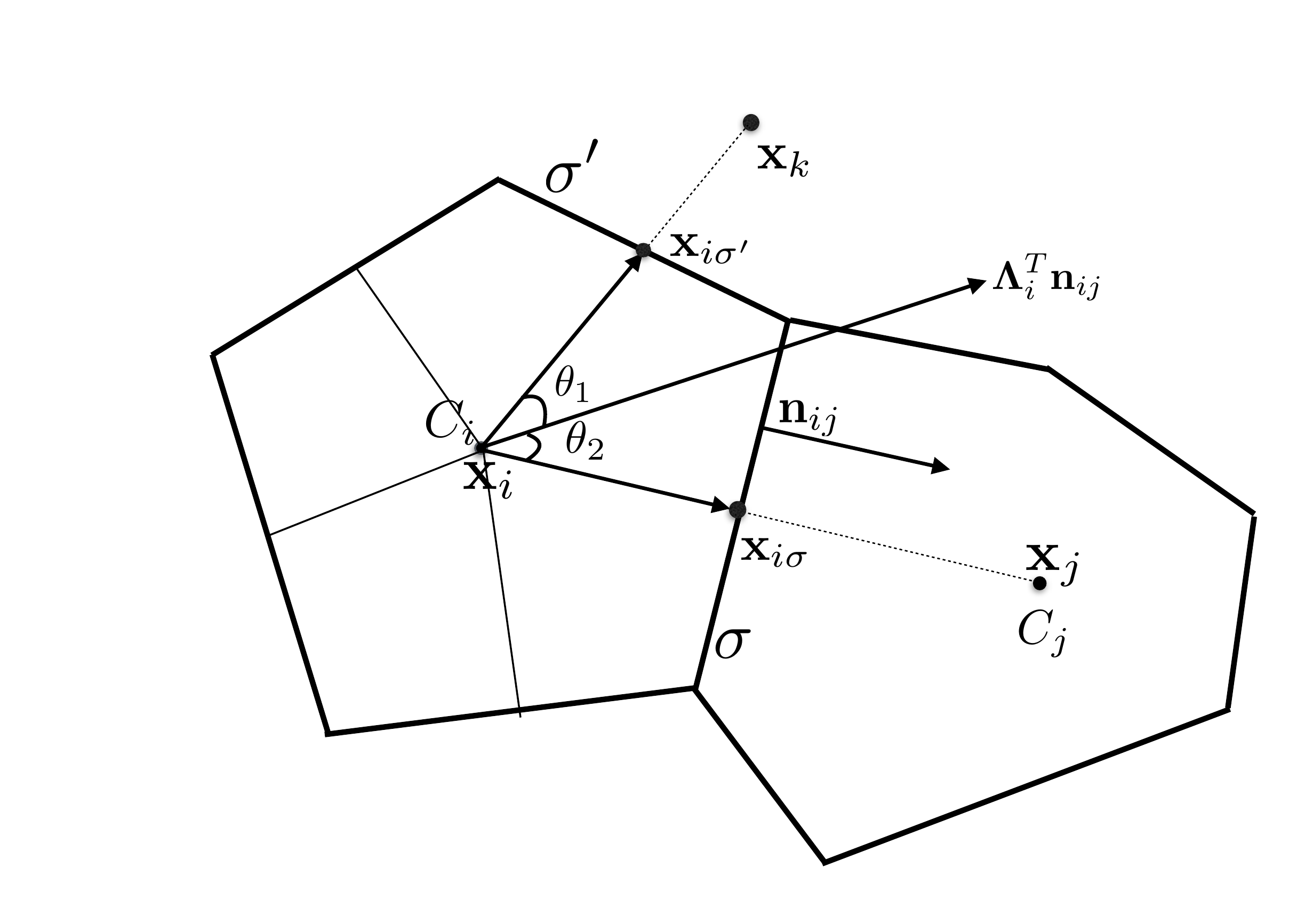}
\caption{A sketch illustrating the calculation of the one-sided
  anisotropic heat flux across the face $\sigma$, which is shared by
  the two Voronoi cells $C_i$ and $C_j$. $\vec{n}_{ij}$ is the unit
  vector perpendicular to $\sigma$, $\vec{x}_i$, $\vec{x}_j$ and
  $\vec{x}_k$ are the centers of cells $C_i$, $C_j$ and $C_k$,
  respectively, $\theta_1$ is the angle between $\vec{x}_{i\sigma}$
  and ${\bf \Lambda}_i^T \vec{n}_{ij}$, $\theta_2$ is the angle between
  $\vec{x}_{i\sigma'}$ and ${\bf \Lambda}_i^T \vec{n}_{ij}$, and
  $\vec{x}_{i\sigma}$ and $\vec{x}_{i\sigma'}$ are the harmonic
  averaging points defined by Eqn.~(\ref{eq:intp}).}
\label{fig:ung}
\end{center}
\end{figure*}

We can also rewrite Equation~(\ref{eq:hfatc}) as 
 \begin{equation}
\vec{j} = -\frac{\kappa}{c_v}[\vec{b}(\vec{b} \cdot \vec{\nabla}u)],
\label{eq:hfatcu}
\end{equation}
by replacing temperature for an ideal gas with the internal energy per
unit mass
\begin{equation}
u = \frac{k_B T}{(\gamma-1)\mu} = c_v T,
\end{equation}
where $\gamma$ is the adiabatic index, and $\mu$ is the mean molecular
weight. The value of $\mu$ depends on the temperature, ionization
state and the composition of the gas. Since we are dealing mainly with
a high temperature, low metallicity plasma of primordial composition
we assume that the gas is fully ionised and consequently set the mean
molecular weight to the constant value $\mu \sim 0.6\, m_p$.  The
final equation for anisotropic thermal conduction then becomes
 \begin{equation}
\frac{\partial u}{\partial t}  =  \frac{1}{c_v \rho}\vec{\nabla}\cdot[\kappa\vec{b}(\vec{b}\cdot\vec{\nabla})u] ,
 \label{dfequ}
\end{equation}
and below we focus on studying numerical solutions schemes for this
partial differential equation.

\section{Methods}
\label{sec:methods}
This section describes the spatial discretisation and time integration 
techniques we employ to solve Eqn.~(\ref{dfequ}).

\subsection{Spatial discretisation}
Consider first the spatial part of Eqn.~(\ref{dfequ}), which takes the
form of a linear operator of the form
\begin{equation}
\mathcal{L}u = \frac{1}{c_v \rho}\vec{\nabla} \cdot [\kappa \vec{b} (\vec{b}\cdot\vec{\nabla})u].
\end{equation}
Since {\sc Arepo} is a finite volume code which divides the
computational domain into a set of control volumes, the above equation
can be recast into the form
\begin{equation}
\mathcal{L}u = \lim_{V\to0} \frac{1}{V}\int_V \frac{1}{c_v \rho}\vec{\nabla} \cdot [\kappa \vec{b} (\vec{b}\cdot\vec{\nabla})u]\, {\rm d}V,
\end{equation}
and by applying Gauss' divergence theorem we get 
 \begin{equation}
 \begin {split}
\mathcal{L}u &=  \lim_{V\to0} \frac{1}{V}\int_{\partial V} \frac{1}{c_v \rho}[ \kappa \vec{b} (\vec{b}\cdot\vec{\nabla})u] \cdot \vec{n} \,{\rm d}A.
\end{split}
\end{equation}
Therefore, the approximated heat flux in a finite-volume scheme 
for each cell $i$ with $N_i$ neighbours is 
 \begin{equation}
 \begin{split}
\mathcal{L}u_i &= \sum_{j = i}^{N_i} {\bf \Lambda}_i  (\vec{\nabla}u)_i \cdot \vec{n}_{ij} ({\rm d}A)_{ij} \\
& =  \sum_{j = i}^{N_i}(\vec{\nabla}u)_i \cdot {\bf \Lambda}_i^T\vec{n}_{ij} ({\rm d}A)_{ij},
\end{split}
\label{eq:disc}
\end{equation}
where 
\begin{equation}
{\bf \Lambda}_i = \frac{\kappa_i}{c_v m_i} 
\begin{cases}
\hfill   \vec{b}_i\otimes\vec{b}_i \hfill & \rm{snisotropic  \ conduction}, \\
\hfill \mathds{1} \hfill & \rm{isotropic \ conduction}, \\
\end{cases}
\end{equation}
and $\vec{n}_{ij}$ is the unit vector perpendicular to the face of
area $({\rm d}A)_{ij}$ shared between cells $C_i$ and $C_j$.

We employ the method described in GW13 for splitting the one-sided
facet heat fluxes into normal and oblique components, followed by a
limiting of the oblique fluxes.  In 2D, the flux splitting is achieved
by decomposing ${\bf \Lambda}_i^T\vec{n}_{ij}$ in terms of two vectors, as
sketched in Fig.~\ref{fig:ung}:
\begin{equation}
\begin{split}
{\bf \Lambda}_i^T \vec{n}_{ij} &= \zeta_{i\sigma} (\vec{x}_{i\sigma} - \vec{x}_i) + \zeta_{i\sigma'} (\vec{x}_{i\sigma'} - \vec{x}_i), \\
\zeta_{i\sigma} &= \frac{||{\bf \Lambda}_i^T\vec{n}_{ij}|| \sin \theta_2}{||\vec{x}_{i\sigma} - \vec{x}_i || \sin( \theta_2 + \theta_1)}, \\
\zeta_{i\sigma'} &= \frac{||{\bf \Lambda}_i^T\vec{n}_{ij}|| \sin \theta_1}{||\vec{x}_{i\sigma'} - \vec{x}_i || \sin( \theta_2 + \theta_1)}, \\
\end{split}
\label{eq:components}
\end{equation}
where $\vec{x}_i$ and $\vec{x}_j$ are the centers of cells $C_i$ and
$C_j$, $\theta_1$ is the angle between $\vec{x}_{i\sigma}$ and
${\bf \Lambda}_i^T \vec{n}_{ij}$, $\theta_2$ is the angle between
$\vec{x}_{i\sigma'}$ and ${\bf \Lambda}_i^T \vec{n}_{ij}$, and the point
$\vec{x}_{i\sigma}$ is the harmonic averaging point defined as
(similar definition applies for $\vec{x}_{i\sigma'}$),
\begin{equation}
\begin{split}
\vec{x}_{i\sigma}({\sigma \in C_i \cap C_j} )= \frac{d_{j\sigma} \lambda_{ij} \vec{x}_i + d_{i\sigma} \lambda_{ji} \vec{x}_j}{d_{j \sigma} \lambda_{ij}  +d_{i\sigma} \lambda_{ji}} .
\end{split}
\label{eq:intp}
\end{equation}
Here, $d_{i\sigma}$ is the orthogonal distance from the center of the cell $C_i$ 
to face $\sigma$, and  $\lambda_{ij} = \vec{n}^T_{ij} {\bf \Lambda}_i \vec{n}_{ij}$.

For a given set of points, a Voronoi tessellation of space consists of
non-overlapping cells around each of the cell generating points such
that each cell contains the region of space closer to it than any of
the other points.  A consequence of this definition is that the cells
are polygons in 2D and polyhedra in 3D, with faces that are
equidistant to the mesh-generating points of each pair of neighboring
cells i.e., $d_{i\sigma} = d_{j\sigma} = d_{ij}/2$. Therefore,
Eqn.~(\ref{eq:intp}) can be rewritten as
\begin{equation}
\vec{x}_{i\sigma} = \omega_{ij} \vec{x}_i + \omega_{ji}   \vec{x}_j,
\end{equation}
where 
\begin{equation}
\omega_{ij} = \frac{\lambda_{ij}}{\lambda_{ij} + \lambda_{ji}}, 
\end{equation}
with $\omega_{ji} = 1 - \omega_{ij}$, and correspondingly, the internal
energy at these points is interpolated as
\begin{equation}
\begin{split}
u({\vec{x}_{i\sigma}}) = u_{\sigma} =  \omega_{ij} u_i + \omega_{ji} u_j.
\end{split}
\label{eq:inpquant}
\end{equation}

Similarly, in 3D, three non-coplanar vectors are needed to express the
anisotropic diffusion vector
\begin{equation}
\begin{split}
{\bf \Lambda}_i^T \vec{n}_{ij} &= \zeta_{i\sigma} (\vec{x}_{i\sigma} - \vec{x}_i) + \zeta_{i\sigma'} (\vec{x}_{i\sigma'} - \vec{x}_i) + \zeta_{i\sigma''} (\vec{x}_{i\sigma''} - \vec{x}_i).
\end{split}
\end{equation}
Hence, we can in general write down ${\bf \Lambda}_i^T \vec{n}_{ij}$ as
\begin{equation} 
\begin{split}
{\bf \Lambda}_i^T \vec{n}_{ij} &= \sum_{\sigma_n, n =1}^{N_{\rm dims}} \zeta_{i\sigma_n}(\vec{x}_{i\sigma_n} - \vec{x}_i) .\\
\end{split}
\label{eq:3D}
\end{equation}
Substituting Eqn.~(\ref{eq:3D}) back into Eqn.~(\ref{eq:disc}), the one-sided flux across the face $\sigma$ from  $C_i$ to $C_j$ is written as,
\begin{equation}
\begin{split}
F_{ij} &= |\sigma| \vec{\nabla}u \cdot \sum_{\sigma_{n}} \zeta_{i\sigma_{n}}(\vec{x}_{i\sigma_{n}} - \vec{x}_i),  \\
&= |\sigma|  \sum_{\sigma_{n}} \zeta_{i\sigma_{n}}(u_{i\sigma_{n}} - u_i).
\end{split}
\end{equation}

Now we have a way of decomposing this flux into normal 
($F^{(1)}$) and oblique ($F^{(2)}$) components,
\begin{equation}
\begin{split}
F_{ij} &= F^{(1)}_{ij} + F^{(2)}_{ij}, \\
F^{(1)}_{ij} &= a_{\sigma \sigma}^i (u_{i\sigma} - u_i),  \\
F^{(2)}_{ij} &= \sum_{\sigma' \neq \sigma} a_{\sigma \sigma'}^i (u_{i\sigma'} - u_i), 
\end{split}
\end{equation}
where $a_{\sigma \sigma'}^i = |\sigma|\zeta_{i\sigma'}$. To make it
locally conservative ($F_{ij} = -F_{ji}$), we rework the net flux in
the following form:
\begin{equation}
\begin{split}
F_{ij}^{\rm{net}} &= \mu_{ij} F^{(1)}_{ij} - \mu_{ji}F^{(1)}_{ji} +
\mu^p_{ij} \left[  1 - {\rm sign} (F^{(2)}_{ij}F^{(2)}_{ji})\right]F^{(2)}_{ij}, \\ 
\rm{where} \ & \mu_{ij} = \frac{|F^{(2)}_{ji}| + \epsilon}{|F^{(2)}_{ij}|+|F^{(2)}_{ji}|+2\epsilon}, \\
\rm{and} \ & \mu^p_{ij} = \frac{|F^{(2)}_{ji}| }{|F^{(2)}_{ij}|+|F^{(2)}_{ji}|+2\epsilon}.
\end{split}
\label{eq:netflux}
\end{equation}
Notice that the oblique flux component depends on the current
estimates of the two one-sided oblique fluxes. 

The salient point is that GW13 prove that this form of the flux is
locally conservative and obeys the DEP under the condition that
$\zeta_{i\sigma} > 0 \ \forall \ [i, \sigma]$ ($\zeta_{i\sigma}$ is
defined by Eqn.~\ref{eq:components}). This condition is automatically
satisfied for a Voronoi mesh as it is convex by construction (see GW13
for more details). A convex polygon is defined as a polygon with all
internal angles less than $180^\circ$. This property implies that no
matter the orientation of ${\bf \Lambda}_i^T\vec{n}_{ij}$, there always
exist $\vec{x}_{i\sigma}$'s such that all $\zeta_{i\sigma}$'s are
positive.

The most computationally intensive part of this calculation is the
estimation of the $\zeta_{i,\sigma}$.  In essence, this involves
determining the triangle within which ${\bf \Lambda}_i^T \vec{n}_{ij}$ lies
in 2D, while in 3D it reduces to determining the tetrahedron which
encompasses this vector. Unfortunately, if a cell $C_i$ has $N_i$
neighbours, a naive implementation of this operation has a complexity
of $\mathcal{O}(N_i^3)$ in 3D, which is highly undesirable. Instead we
make use of the fact that {\sc Arepo} first creates a Delaunay
tessellation and then obtains the Voronoi tessellation from it. The
Delaunay tessellation is formed by triangles/tetrahedra that do not
contain any of the points inside their circumspheres (see
\citealt{Springel2010} for more details).  Therefore, in order to
compute the $\zeta_{i,\sigma}$, we just determine in which of the
Delaunay triangles/tetrahedra connected to the cell $C_i$ the vector
${\bf \Lambda}_i^T \vec{n}_{ij}$ lies. This reduces the complexity of the
algorithm to $\mathcal{O}(N_{\rm tetra})$, where $N_{\rm tetra}$ are all
the points which have a Delaunay connection to $C_i$ (these can also
be points where the area of the face between the points is zero).

In passing we note that the dynamic movement of the mesh in {\sc
  Arepo} does not affect our basic implementation.  Mesh motion is
completely taken care of by the hydro solver; the diffusion solver
only sees a static mesh since we couple it to the rest of the dynamics
through operator splitting.

\subsection{Time integration}
\label{sec:ti}

The simplest way to perform the time integration is to do it
explicitly. However, numerical stability requires that the timestep is
limited by the von Neumann stability condition
$\Delta t \leq \eta_{N}(\Delta x)^2/\chi$ where $\Delta x$ is the cell
width, $\chi = \kappa/(c_v\rho)$ and $\eta_{N} < 1$ is the von Neumann
stability coefficient which we usually set to $0.2$. This timestep
constraint becomes rather severe, especially at high resolution (given
its quadratic dependence on $\Delta x$) and in high temperature
plasmas.  Therefore, an implicit method that places no stability limit
on the diffusion timestep is highly desirable.

The easiest implicit time discretisation is the first order backwards
Euler method,
\begin{equation}
\frac{u_i^{t+\Delta t} - u_i^t}{\Delta t} = \sum_{j=i}^{N_i} A_{ij} (u_j^{t + \Delta t} - u_i^{t + \Delta t}),
\label{eq:be}
\end{equation}
where $A_{ij}$ is the coefficient matrix defined as
\begin{equation}
F_i = \sum_{j}A_{ij} (u) (u_j - u_i).
\end{equation}
$A_{ij}$ can be split into normal ($\alpha_{ij}$) and oblique ($\beta_{ij}$) components,
\begin{equation}
\begin{split}
A_{ij} &= \alpha_{ij}+ \beta_{ij}(u), \\
\end{split}
\end{equation}
which in turn are defined as
\begin{equation}
\begin{split}
\alpha_{ij} &= \mu_{ij}a_{\sigma\sigma}^i\omega_{ji} + \mu_{ji}a_{\sigma\sigma}^j\omega_{ij}, \\
\beta_{ij}(u) &= \omega_{ji}\sum_{k, \sigma' \in C_i \cap C_k}
\mu_{ik}^p \  [1-{\rm sign}(F^{(2)}_{ik}(u)F^{(2)}_{ki}(u)) ]\ a_{\sigma'\sigma}^i.
\end{split}
\end{equation}
Equation~(\ref{eq:be}) can be rewritten as
\begin{equation}
\begin{split}
u_i^{t+\Delta t} - \Delta t \sum_{j=i}^{N_i} A_{ij} (u_j^{t + \Delta t} - u_i^{t + \Delta t})  = u_i^t . 
\end{split}
\end{equation}
Focussing on the left-hand-side of this equation, we obtain
\begin{equation}
\begin{split}
&{\rm L.H.S} = u_i^{t + \Delta t}  + \Delta t \sum_{k=i}^{N_i} A_{ik} u_i^{t + \Delta t} - \Delta t \sum_{j=i}^{N_i} A_{ij} u_j^{t + \Delta t} \\
&= \sum_{j=i}^{N_i} \left( \delta_{ij} u_j^{t + \Delta t} + \Delta t \sum_{k=i}^{N_i} A_{ik} \delta_{ik} u_k^{t + \Delta t} \right) - \Delta t \sum_{j=i}^{N_i} A_{ij} u_{j}^{t + \Delta t}. \\
\end{split}
\end{equation}
Therefore the equation we have to solve is given by
\begin{equation}
\begin{split}
\sum_{j=i}^{N_i} \left( \delta_{ij} \left(1+\Delta t \sum_{k=i}^{N_i} A_{ik} \right) - \Delta t A_{ij} \right) u_j^{t + \Delta t} = u_i^t , \\
\end{split}
\label{eq:iti}
\end{equation}
which is of the generic form
\begin{equation}
\vec{A}(\vec{U})\,\vec{U} = \vec{U}_0 ,
\label{eq:Picard}
\end{equation}
where $\vec{U}_0 = \{u_i^t\}$ is the internal energy vector at the
present time. The dependence of the coefficient matrix $\vec{A}$ on
the internal energy vector makes the the system non-linear.

This nonlinear system can be solved by Picard's iterative method
\citep{Picard}.  This approach linearizes the non-linear system by
estimating the value of the coefficient matrix from the values of the
internal energy obtained from the previous iteration. The algorithm to solve the
full non-linear system is then given by
\begin{enumerate}
\item Solve the linear system
  $\vec{A}(\vec{U}^{n-1}) \vec{U}_*^n = \vec{U}_0$ with respect to
  $\vec{U}_*^n$.
\item Relax the solution $\vec{U}^n = \mu^n \vec{U}_*^n + (1 - \mu^n)\vec{U}^{n-1}$, where $\mu^n$ is relaxation parameter given by Eqn.~(\ref{eq:umc}).
\item Break if $\frac{||\vec{A}(\vec{U}^n)\vec{U}^n - \vec{U}_0||}{||\vec{A}(\vec{U}_0)\vec{U}_0 - \vec{U}_0||} <
  10^{-3} $ or $\frac{||\vec{U}^n - \vec{U}^{n-1}||}{||\vec{U}_0||} < 10^{-6}$. 
\item Else set $\vec{U}^{n-1} = \vec{U}^n$, and repeat.
\end{enumerate}

We use the {\sc hypre}
library\footnote{\href{http://acts.nersc.gov/hypre}{http://acts.nersc.gov/hypre}}
to solve the linear system in step (i). {\sc hypre} is a library for
solving large, sparse linear systems of equations on massively
parallel computers \citep{HYPRE}. In particular, the linear system is
solved using the generalised minimal residual (GMRES) iterative method \citep{Saad1986}. Additionally an algebraic multi-grid preconditioner \citep{Henson2002} is used in order to achieve faster convergence.
The tolerance limit for iteratively solving the linear system is set
to $\epsilon_{\rm lin} = 10^{-8}$.

A Picard iteration consists of solving
$\vec{A}(\vec{U}_-)\vec{U} = \vec{B}$ until convergence.  This has
several drawbacks in practice. If successive correction vectors,
$\vec{C}^{n-1} = \vec{U}^n - \vec{U}^{n-1}$ and
$\vec{C}^n = \vec{U}^{n+1} - \vec{U}^n$, are in roughly the same
direction, the scheme is undershooting. This means $\vec{C}^{n-1}$
could have been larger, reducing the number of iterations. This can in
principle be improved by overrelaxation, i.e.~taking instead of
$\vec{U}^n$ a larger step $\mu \vec{U}^n$ with $\mu > 1$. On the other
hand, if $\vec{C}^n$ is roughly in the opposite direction to
$\vec{C}^{n-1}$, the scheme is overshooting the solution.  In the
worst case, this can lead to an endless sequence of correction vectors
oscillating around the actual solution. Overshooting can be
remedied by underrelaxation, i.e.~taking a smaller step
$\mu \vec{U}^n$ with $0 < \mu < 1$.  

We use a variant of the unstable manifold corrector scheme outlined in
\citep{DeSmedt2010} to adaptively under- or overrelax the solution (as
in step (iii) of the non-linear implicit algorithm). Let us define the
angle ($\theta$) between the correction vectors as
\begin{equation}
  \theta^n = { \rm cos}^{-1} \left(  \frac{(\vec{C}_{*}^{n-1})^T \cdot \vec{C}^{n-2}}{||\vec{C}_{*}^{n-1}|| \cdot ||\vec{C}^{n-2}||}  \right),
\end{equation}
and then over- or underrelax the solution according to the value of $\theta$,
\begin{equation}
\mu^n = 
\begin{cases}
\hfill 1.5 \hfill & \theta^n \leq \frac{\pi}{8} \\
\hfill 1.0 \hfill & \frac{\pi}{8}< \theta^n \leq \frac{3\pi}{4} \\
\hfill 0.5 \hfill & \frac{3\pi}{4}< \theta^n \leq \frac{9\pi}{10} \\
\hfill 0.3 \hfill & \theta^n > \frac{9\pi}{10}. \\
\end{cases}
\label{eq:umc}
\end{equation}

  \begin{figure*}
\begin{center}
\includegraphics[width=1.\textwidth]{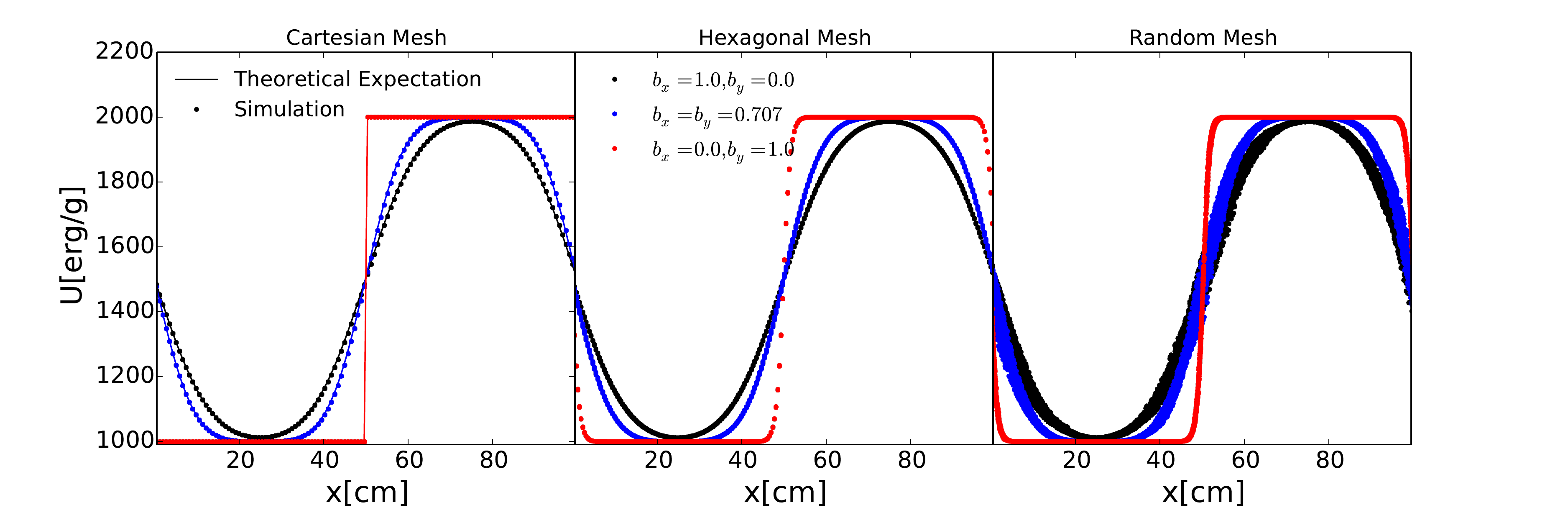}
\caption{Diffusion of a temperature step function in the presence of a
  magnetic field in the direction of the temperature gradient (black
  curves), inclined by $45 ^ \circ $ (blue curves) relative to the
  temperature gradient, and perpendicular to the temperature gradient
  (red curves) at $t=5\,{\rm s}$.  The test is performed for three
  different configurations of the mesh, regular Cartesian (left
  panel), regular hexagonal (middle panel) and a random mesh (right
  panel). The resolution is $100^2$ particles within the domain. The
  points are the simulations results and the solid curves are the
  theoretical expectations.}
\label{fig:step}
\end{center}
\end{figure*}

This fully implicit scheme is unconditionally stable. However, it is
also computationally expensive because we need to perform non-linear
iterations on top of the linear iterations at each step. In our test
problems, presented in Section~\ref{tests}, an average of 10-15
nonlinear iterations were required per timestep. This reflects the
fact that in order to achieve a significant speedup in the
calculation, one needs to set $\eta_{N} \ge 100$.  Therefore, this
method is only viable if the magnetic fields are slowly varying, over
characteristic timescales of order $ \sim 100$ times the conduction
timescale.

Therefore, for problems where the accuracy over a short period of the
magnetic field is important (for example see Section~\ref{sec:intb})
we implement a semi-implicit algorithm following the method outlined
in~\citet{Sharma2011}.  Here the terms of the coefficient matrix that
depend on the internal energy are integrated explicitly, while the
other terms are integrated implicitly.  Accordingly,
Eqn.~(\ref{eq:iti}) changes form and becomes
\begin{equation}
\begin{split}
\sum_{j=i}^{N_i} \left( \delta_{ij} \left(1+\Delta t \sum_{k=i}^{N_i} \alpha_{ik} \right) - \Delta t \alpha_{ij} \right) u_j^{t + \Delta t} = \\ u_i^t + \Delta t \sum_{l=1}^{N_i} F^{(2)}_{il}(u_l^t, u_i^t). \\
\end{split}
\end{equation}
In this form, the coefficient matrix is independent of the current
value of the internal energy, and therefore the system is linear. As
with the fully implicit scheme, this linear system is solved using the
\textsc{hypre} library with the GMRES routine and a multigrid
preconditioner. We numerically verify that this method is stable up to
$\eta_{N} = 4$, which is still an improvement by a factor of
$\sim 10-20$ over the explicit scheme. As mentioned in
\citet{Sharma2011}, the explicit treatment of the oblique flux terms
means that this scheme is not strictly obeying the DEP, but the
deviations from it are very small and the corresponding oscillations
are damped with time. This method is extremely fast as it requires
only the solution of a linear system per timestep.

One drawback of the implicit schemes is that it is not directly
compatible with the individual time-stepping method used in {\sc
  Arepo} for advancing ordinary hydrodynamics. We presently overcome
this by solving the conduction equation only on global synchronization
timesteps of the code. This is a viable approach because the fully
implicit scheme is unconditionally stable and does not pose strong
constraints on the permissible timestep.  However, the timestep
restriction of the semi-implicit method implies that when this method
is used we need to restrict the global timestep in the simulation to
$\eta_{N}$ times the minimum of the diffusion step size of all cells
in the domain

\section{Numerical Tests}
\label{tests}
In this section we present the results from various numerical tests of our  anisotropic conduction implementation.  The relevant hydrodynamics
equations are evolved using the moving mesh finite volume scheme outlined in 
\citet{Springel2010}, with the improved time integration
and gradient estimation techniques described in \citet{Pakmor2016}.
The magnetic fields which decide the direction of diffusion are
evolved using the ideal MHD module outlined in \citet{Pakmor2013}
based on the 8-wave formalism to control divergence errors
\citep{Powell1999}.

Sections~\ref{sec:dsf}, \ref{sec:ring} and \ref{sec:sovinec} are
intended to test the accuracy and the stability of the anisotropic
conduction implementation only, without evolving the hydrodynamical
equations. Therefore, in all these tests, the cells do not move and
only interact with each other through heat conduction (i.e.~the gas
dynamics is not explicitly followed and the gas velocity is fixed to
zero at all times). This also ensures that the magnetic fields are
constant with time.

In Sections~\ref{sec:bw} and \ref{sec:intb}, we additionally test the
accuracy of the anisotropic conduction scheme when hydrodynamics and
heat conduction are coupled.  In these tests, we initially start with
a regular Cartesian grid, which is then allowed to move and distort
according to the local fluid motion \citep{Springel2010}. In addition,
the mesh is regularized where needed using the scheme outlined
in~\citet{Vogelsberger2012}.  The time integration in all these tests
is performed with the semi-implicit scheme outlined in
Section~\ref{sec:ti}, using $\eta_{N}=4$.

\subsection{Diffusion of a step function} 
\label{sec:dsf}

As a first test of our implementation, we study the diffusion of a
temperature step function. We perform 2D simulations with a domain
size of $(100 \, {\rm cm})^2$ sampled with $(100)^2$ cells. The
internal energy (in erg/g) of a given cell is set to
\begin{equation}
u(x,y) = 
\begin{cases}
\hfill 1000 \hfill & x\leq50 \\
\hfill 2000 \hfill & x > 50.
\end{cases}
\end{equation}
The density is kept constant at $\rho = 1\, \rm{g \, cm^{-3}}$
throughout the computational domain. We set the diffusivity to be
\begin{equation}
\chi = \frac{\kappa}{c_v \rho} = 10 \ \rm{cm^2\,s^{-1}},
\end{equation}
and adopt periodic boundary conditions at the domain edges.
The analytic solution for this problem under periodic boundary
conditions is given by 
  \begin{equation}
  \begin{split}
  U(x,t) &= u_0  + \frac{\Delta u}{2} \sum_{x_i, i=1}^3 \left[
    (-1)^i{\rm erf}\left( \frac{x - x_i}{\sqrt{4\chi  t b_x^2}}
    \right)\right], 
  \end{split}
  \label{eq:step_diff}
\end{equation}
with $x_1 = 0$, $x_2 = 50$, $x_3 = 100$, $u_0=1500$ and
$\Delta u =1000$.

We perform this test with three different kinds of meshes, Cartesian,
hexagonal and random. The random mesh is created by choosing points
inside the simulation domain whose coordinates are a pair of random
deviates extracted from a uniform probability distribution in the
interval bounded by the domain and letting {\sc Arepo} construct a
Voronoi mesh out of these random points.  Fig.~\ref{fig:step} shows
the results of the temperature step function test on the three
different meshes at time $t=5\, s$. The red points show the internal
energy evolution when the magnetic field is perpendicular to the
temperature gradient ($\vec{b} = \{0,1\}$), blue points when the
magnetic field and the temperature gradient are are misaligned by
$\pi/4$ ($\vec{b}=1/\sqrt{2}\{1,1\}$) and the black points when the
magnetic field direction and the temperature gradient is aligned
($\vec{b}=\{1,0\}$). The left panel shows the result for the Cartesian
mesh, the middle panel for the hexagonal mesh and the right panel for
the random mesh.

It is evident that the simulation results agree well with the analytic
solution (solid curves). We have also verified that our implementation
obeys the DEP in this test. Even when the magnetic field is
perpendicular to the direction of the temperature gradient, the
algorithm performs remarkably well and does not give rise to unstable
oscillations. \citet{Arth2014}, for example, had to add an isotropic
component to their anisotropic diffusion tensor in order to stop
unphysical oscillations. While we do have some amount of perpendicular
numerical diffusion, this appears to be very small, something we will
further quantify in Section~\ref{sec:sovinec}.

Another point to notice is that although the random mesh simulations
do well on average, i.e., the mean temperature of the cell at the
particular value of $x$ matches the analytical expectation well, there
is some amount of scatter around the mean. We argue that this scatter
does not arise because of an incorrect calculation of the heat fluxes
on a random mesh, but because the boundaries of the random mesh are
not regularly spaced in $x$ (i.e.~along the direction of the
temperature gradient).  The difference in sizes of the cells in a
random mesh makes the conduction front irregular, giving rise to the
observed temperature scatter.

\begin{figure*}
\begin{center}
\includegraphics[scale=0.70]{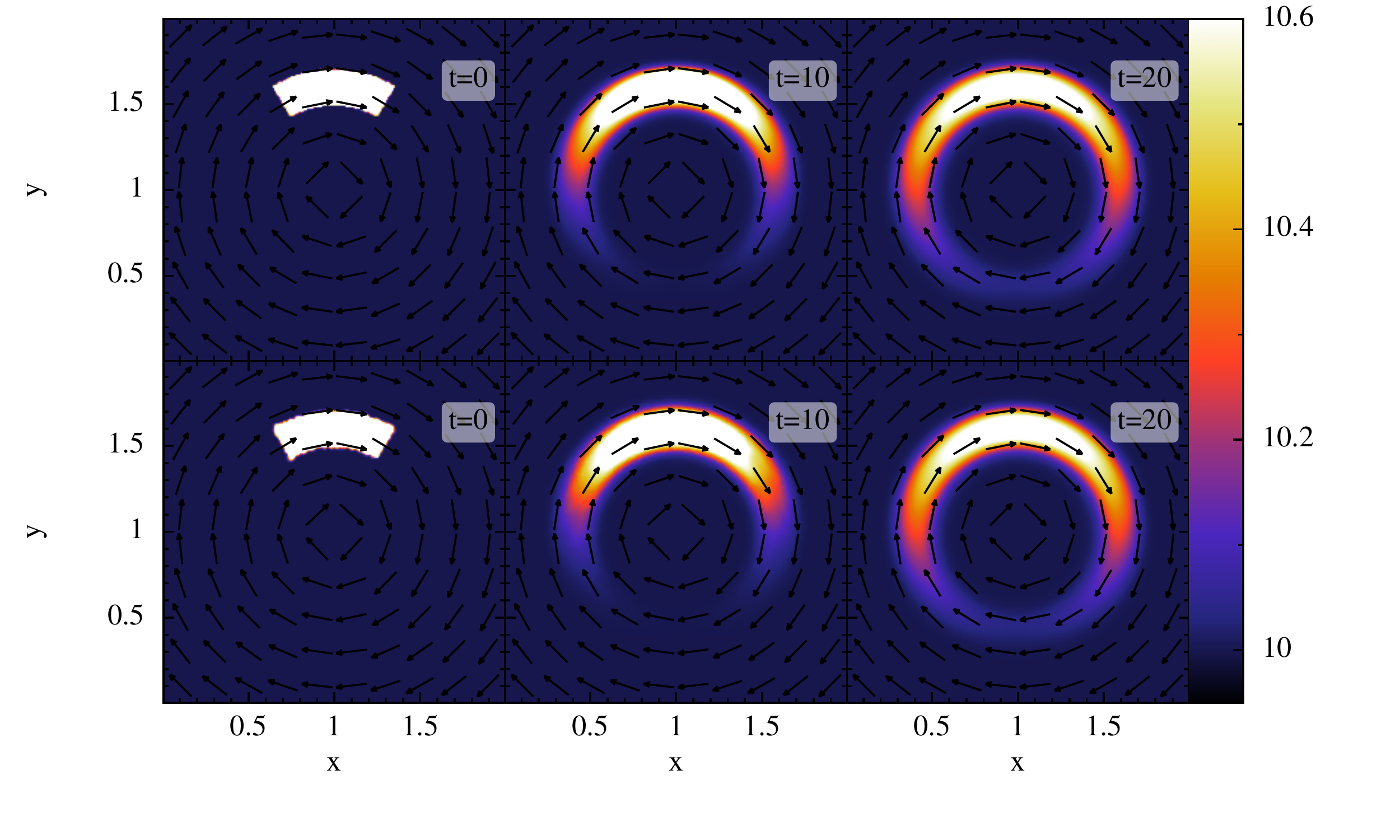}
\caption{Time evolution of a hot patch of gas under a circular
  magnetic field in 2D.  The test is performed for both a regular
  Cartesian (top panel) and a random mesh (bottom panel). The
  resolution is $400^2$ particles within the domain. The arrows
  represent the orientation of the magnetic field and the colour bar
  shows the internal energy of the gas in ${\rm erg\,g^{-1}}$. This
  plot was created using the interactive visualization tool {\sc
    Splash} \citep{Price2007}.}
\label{fig:ring}
\end{center}
\end{figure*}

\subsection{Diffusion of a hot patch of gas in a circular magnetic field}
\label{sec:ring}

We investigate the circular diffusion problem as outlined in
\citet{Parrish2005} and \citet{Sharma2007}. A hot patch of gas
surrounded by a cooler background is allowed to diffuse in the
presence of a circular magnetic field. The temperature drops
discontinuously across the patch boundary. The computational domain is
of size ($2 \ \rm{cm}$)$^2$ with periodic boundary conditions.

We perform this test for a regular Cartesian grid and a random mesh
(as in Section~\ref{sec:dsf}). The regular Cartesian mesh describes
the behaviour of the code in the best case scenario where the
regularity of the mesh inherently reduces the errors, whereas the
random mesh corresponds to the worst case scenario where in certain
situations the mesh can be stretched and distorted by large amounts
due to fluid motion. The behaviour of the mesh and hence the degree of
mesh related errors can be controlled by regularizing the mesh, using
schemes such as the ones described in \citet{Springel2010},
\citet{Vogelsberger2012} and \citet{Mocz2015}.

The initial internal energy (in $\rm{erg/g}$) distribution is given by
\begin{equation}
u(x,y) = 
\begin{cases}
\hfill 12 \hfill & \mbox{for} \ 0.5<r<0.7 \ \rm{and} \ abs(\theta)< \pi/6 , \\
\hfill 10 \hfill \ &  \rm{otherwise},
\end{cases}
\end{equation}
where $r = \sqrt{(x-1)^2 + (y-1)^2}$ and
$\theta = \tan^{-1}[(y-1)/(x-1)]$. The magnetic fields are circular
and centered on [1,1]. The density is set to unity. The diffusivity
$\chi$ is set to $0.01 \ \rm{cm^2\,s^{-1}}$. The simulation is run
until time $t=200\,{\rm s}$. There is no explicit perpendicular
diffusion coefficient, so ideally the internal energy outside the ring
($r\leq0.5$ and $r\geq0.7$) stays always $10$. The initial hot gas should
only diffuse along the azimuthal magnetic field until at the end it is
evenly distributed in a ring.

\begin{figure}
\begin{center}
\includegraphics[scale=0.45]{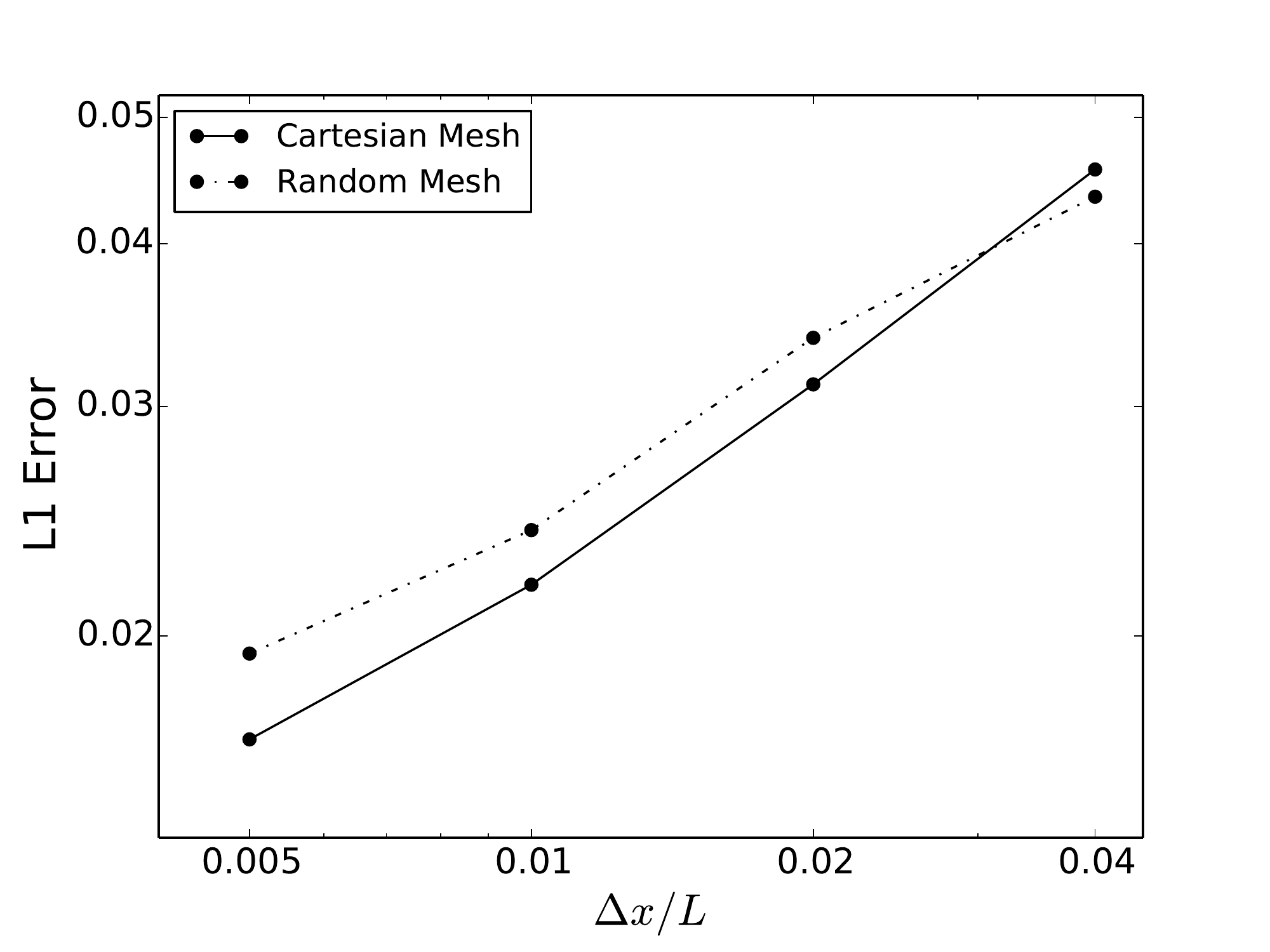}
\caption{L1 error, at $t=10\,{\rm s}$, as a function of resolution for
  a Cartesian (solid line) and a random mesh (dot dashed line) for
  simulations of the diffusion of a hot patch of gas in circular
  magnetic fields (Section \ref{sec:ring}). The convergence order for
  the Cartesian mesh is $0.49$, and $0.41$ for the random meshes.}
\label{fig:conv}
\end{center}
\end{figure}

We note that in the coordinate defined as $s=r\theta$, the problem
reduces to the diffusion of a double step function. This is valid
while no material has diffused more than halfway across the
ring. Therefore, the analytic solution is of the same form as
Eqn.~\ref{eq:step_diff} in the co-ordinate $s$ with $b_x =1$.
Fig.~\ref{fig:ring} shows the internal energy distribution at times
$t=0\,{\rm s}$ (left panels), $t=10\,{\rm s}$ (middle panels) and
$t=20\,{\rm s}$ (right panels) for both a Cartesian (top panels) and
random (bottom panels) grid simulations.  The evolution of the
internal energy is quite similar and is independent of the mesh
used. We further quantify this by looking at the L1 error in the
simulations. Figure~\ref{fig:conv} plots the L1 error at
$t=10\,{\rm s}$ in the simulation as a function of the resolution for
both Cartesian (solid curve) and random (dot-dashed curve) meshes.

The convergence order for the Cartesian mesh is about $0.49$ and for
the random mesh about $0.41$. This convergence order is smaller than
the limited symmetric method, but similar to the values obtained by
other slope limited schemes \citep{Sharma2007}. A better interpolation
scheme and a better non-linear flux limiter that is both conservative
and DEP-preserving can potentially improve the convergence order of
our scheme. However, it is quite a challenge to achieve this,
especially for unstructured meshes.

Perpendicular diffusion is an unwanted side effect of the non-linear
flux limited scheme.  Although the limited fluxes are more diffusive,
we numerically verify that they never undershoot below the minimum or
overshoot above the maximum temperature. The limiter hence ensures the
stability of the scheme but reduces the accuracy.

\subsection{Sovinec test}
\label{sec:sovinec}

The limiting of the oblique fluxes results in higher perpendicular
numerical diffusion. It is quite important to quantify this artificial
numerical diffusivity in order to understand the limitations of our
implementation. We estimate the numerical diffusivity by performing
the experiment suggested in \citet{Sovinec2004}.  We perform this test
for both Cartesian and random meshes (as in Section~\ref{sec:dsf}) in
2D.

Let us consider the full heat conduction equation (i.e.~a
generalization of Eqn.~\ref{dfequ}),
\begin{equation}
\frac{\partial u}{\partial t}  =
\vec{\nabla}\cdot[\kappa_{||}\vec{b}(\vec{b}\cdot\vec{\nabla})u] +
\vec{\nabla}\cdot[(\kappa_{\perp} + \kappa_{\rm num})(\mathds{1} - (\vec{b} \otimes \vec{b}))\vec{\nabla}u] + Q.
\label{eq:sf}
\end{equation}
where $Q$ is a source function of the form
$Q = Q_0 \cos(k x) \cos(k y)$, and we set $\rho c_v = 1$. We further
assume a fixed magnetic field such that $\vec{B}\cdot\vec{\nabla}u= 0$
and also set $\kappa_{\perp} =0$. We use non-periodic boundary
conditions with the temperature at the boundary fixed to
$T_{\rm bound} = 0$, which means that the boundaries act as a heat
sink. Then the steady state solution of Eqn.~(\ref{eq:sf}) at the
center of the domain is $u(0,0) = Q_o/(2k^2 \kappa_{\rm
  num})$.
Therefore the perpendicular numerical diffusivity can be easily
deduced from the value of the central internal energy.

\begin{figure}
\begin{center}
\includegraphics[scale=0.45]{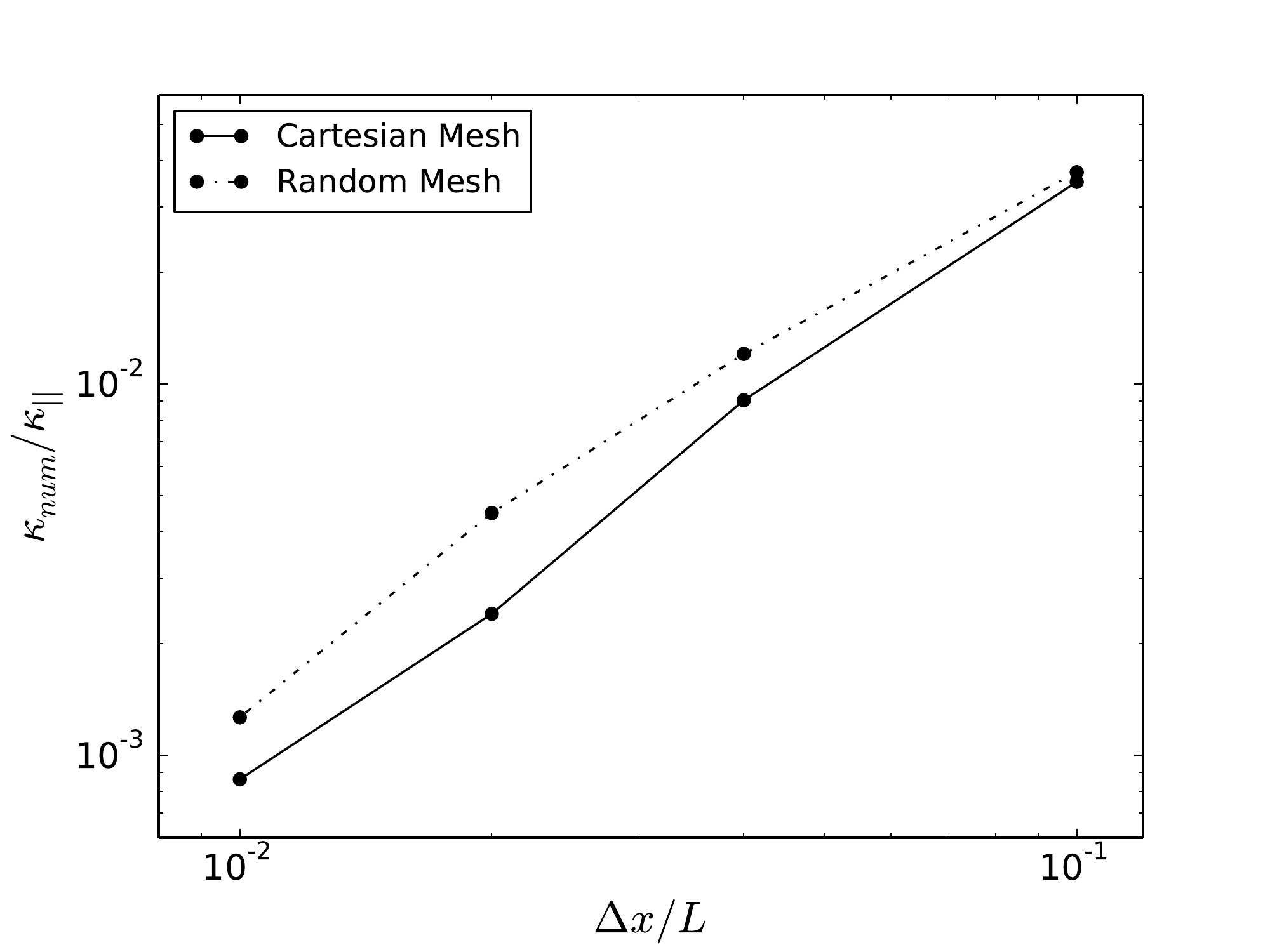}
\caption{Relative numerical perpendicular diffusion coefficient as a
  function of resolution as obtained from the Sovinec test
  (Section~\ref{sec:sovinec}). The convergence order is about $1.63$
  for Cartesian meshes (solid line) and $1.31$ for random meshes
  (dot-dashed line).}
\label{fig:sovinec}
\end{center}
\end{figure}

We simulate this configuration in a domain of size
$[-0.5,0.5]\times[-0,5,0.5]$ with density $\rho = 1$, $k=\pi$ and
$Q_0 = 2\pi^2$. The initial temperature is
$T(x,y) = \cos(\pi x) \cos(\pi y)$ and the magnetic field components
are set to $B_x = \cos(\pi x) \sin(\pi y)$ and
$B_y = -\sin(\pi x) \cos(\pi y)$, so that the value of the numerical
perpendicular diffusivity can be estimated from
$\kappa_{\rm num} = 1/u(0,0)$ once the system reaches the steady
state.  We then measure the ratio $\kappa_{||}/\kappa_{\rm num}$ as a
function of resolution and plot the results is
Fig.~\ref{fig:sovinec}. We see that even at really low resolution the
value of the perpendicular numerical diffusivity is just about $3\%$
the value of $\kappa_{||}$. This is comparable to the numerical
diffusivity obtained from slope limited methods such as in
\citet{Sharma2007}. We also note that the convergence order in
Cartesian meshes is about $-1.63$, again in agreement with previous
methods. The convergence order for random meshes is about $-1.31$,
which is lower than the one determined for Cartesian meshes. This is
mainly caused by the different cell sizes (and therefore different
\textit{local} resolutions) in different parts of the domain.

\subsection{Point explosion with heat conduction}
\label{sec:bw}

We next test our conduction implementation for the Sedov-Taylor blast
wave problem. This problem is a good test to determine the accuracy of
coupling two distinct physical processes: hydrodynamics and diffusion.
We simulate three different configurations of the blast wave problem, the
classical adiabatic blast wave test, the blast wave test with
isotropic conduction, and the blast wave test with anisotropic
conduction in 3D. We simulate a domain of size $100$ pc on a side with
$64^3$ resolution elements to begin with.

\begin{figure*}
\begin{center}
\includegraphics[scale=0.80]{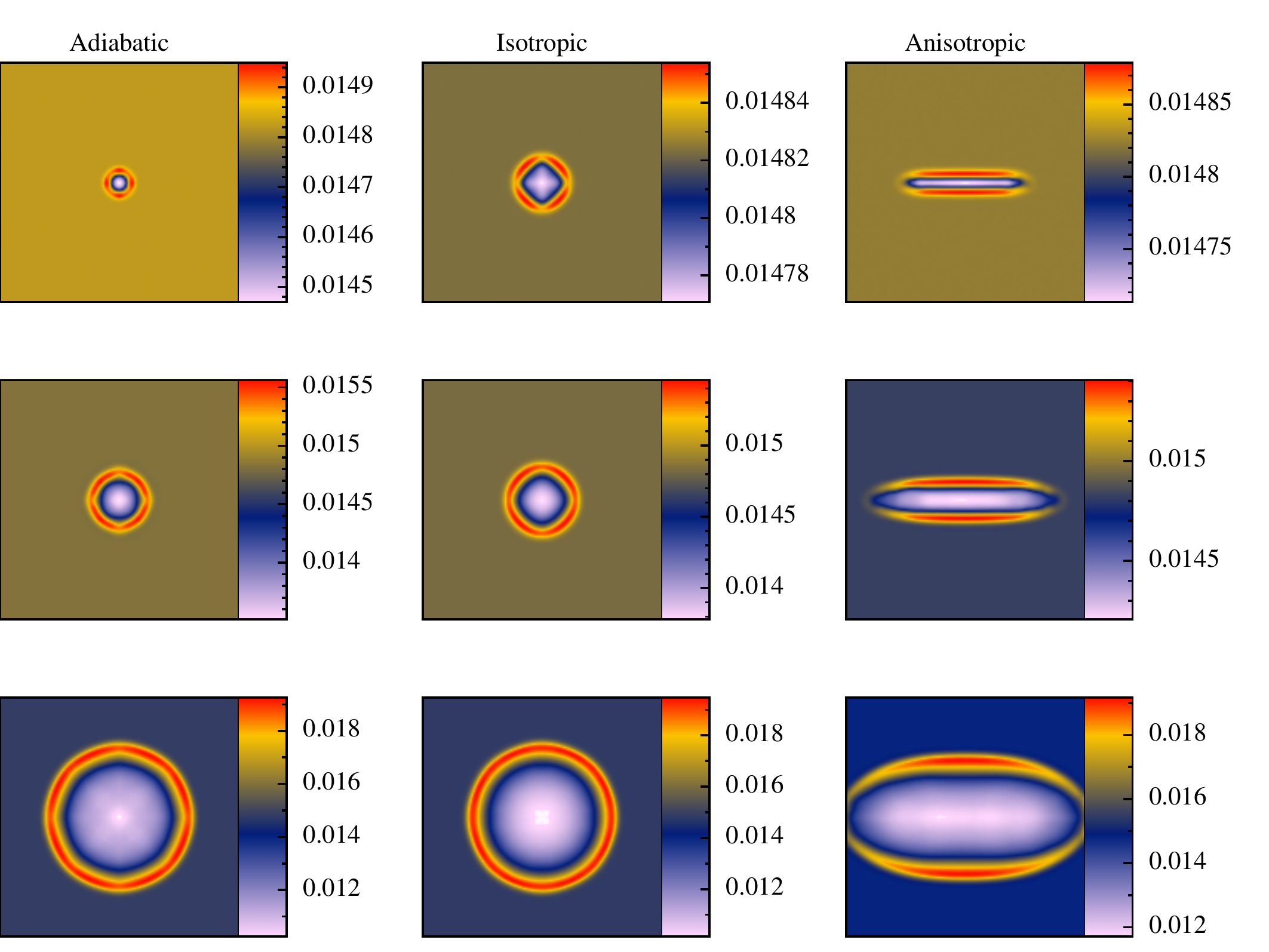}
\caption{Projected gas density maps in the $xy$-plane of an adiabatic
  point explosion (left column), and point explosions with isotropic
  (middle column) and anisotropic (right column) conduction at 1 (top
  row), 10 (middle row) and 100 (bottom row) kyrs. The magnetic field
  strength is $1 {\rm \mu G}$, pointing in the positive
  $x$-direction. The size of each panel is (100 pc)$^2$. The
  resolution is $64^3$ particles within the domain. This plot was
  created using the interactive visualization tool {\sc Splash}.}
\label{fig:blast_wave}
\end{center}
\end{figure*}

We inject $E_0 = 3.33 \times 10^{50} \ \rm{erg}$ of energy into the
central 8 cells.  The blast wave expands into a uniform medium of
density $\rho_0 = 1 \ \rm{cm}^{-3}$ and temperature
$T = 10^4 \ {\rm K}$. A uniform magnetic field pointing in the positive
$x$-direction of strength $1\, {\rm \mu G}$ is added to each gas
cell. This low value of the magnetic field ensures that it is not
important for the dynamics of the explosion. We hence expect the blast
wave to expand spherically without any hindrance from the magnetic
field. All three simulations are performed for a total of $10^5$ yr.
The adiabatic index is set to $\gamma = 5/3$. For this test problem,
we do not include the effect of saturation when calculating the
conduction heat flux and we set the value of the conduction
coefficient $\kappa$ to the full Spitzer value (Eqn.~\ref{eq:ksp}).

Fig.~\ref{fig:blast_wave} shows the resulting density maps of the
adiabatic (left column), isotropic conduction (middle column) and
anisotropic conduction (right column) simulations at 1 (top row), 10
(middle row) and 100 (bottom row) kyrs. We can see right away that the
results are very different. Qualitatively, the conduction front
outpaces the shock front during the initial phase of expansion ($<10$
kyrs) in the isotropic conduction simulation, while anisotropic
conduction seems to accelerate the shock in the direction of the
magnetic field while the shock front in the perpendicular direction
lags behind. At late times (100 kyrs), the shock front in the
isotropic and the adiabatic cases appears to be in the same
position. The perpendicular shock front in the anisotropic conduction
simulation still lags behind.

Figures~\ref{fig:bw3} and \ref{fig:bw80} show the results in a more
quantitative manner at two different times, $3$ kyr and $80$ kyr,
respectively. The solid black lines correspond to the classical
adiabatic solutions for the pressure, density and temperature
structure in the shock. As we can see clearly, the adiabatic
simulation (left column) accurately traces the theoretical
expectations \citep{Sedov1959} for the pressure (top row), density
(middle row) and temperature (bottom) structure behind the shock. In
the following subsections, we describe in more detail the structure of
the solution when isotropic and anisotropic thermal conduction are
included in the calculations.

\subsubsection{Point explosion with isotropic heat conduction}

The evolution of the point explosion with heat conduction was
considered in detail by \citet{RMTV1991}. They obtain self-similar
solutions under the assumption that the ambient gas density decays
with a given power of the radius.  However, if the density is
constant, then the solution is no longer self-similar. However, the
solution can be obtained by analysing the corresponding pure
hydrodynamics and pure diffusion problems, each of which has a
similarity solution.

\begin {figure}
\begin{center}
\includegraphics[scale=0.35]{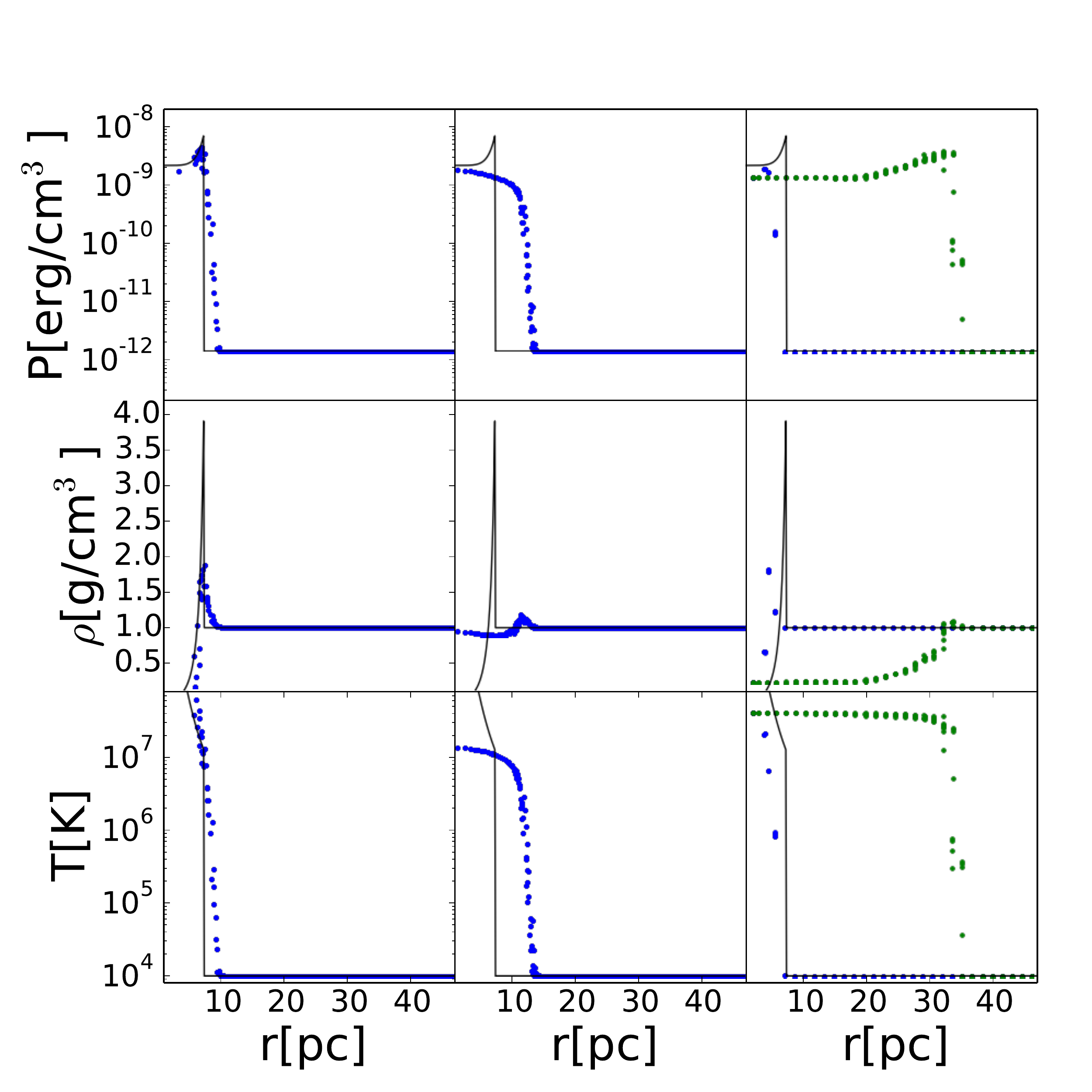}
\caption{Pressure (top row), density (middle row) and temperature
  (bottom row) profiles in the point explosion with adiabatic (left
  column), isotropic (middle column) and anisotropic conduction (right
  column) at $t=3$ kyr. The solid lines plot the analytic solution of
  the classical adiabatic Sedov-Taylor problem. The blue points are
  the simulation results. For anisotropic conduction, we plot two
  profiles, one along the magnetic field direction (green points) and
  the other perpendicular to it (blue points).  }
\label{fig:bw3}
\end{center}
\end{figure}

\begin{figure}
\begin{center}
\includegraphics[scale=0.35]{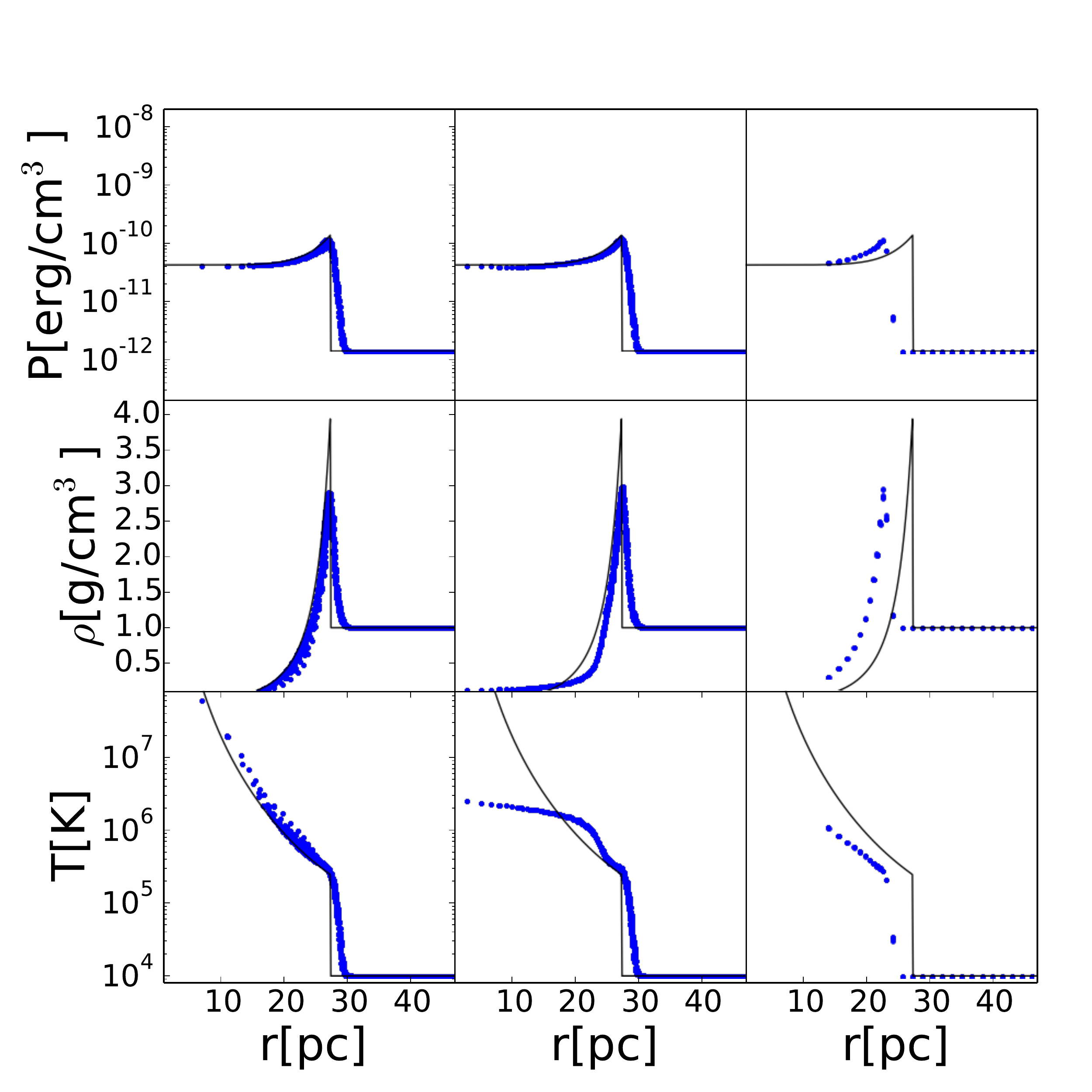}
\caption{Same as Fig.~\ref{fig:bw3} but for $t=80\,{\rm kyr}$. The
  anisotropic conduction plots do not have any green simulation points
  because the shock front in the direction of the magnetic field has
  progressed beyond the computational domain.}
\label{fig:bw80}
\end{center}
\end{figure}

The position $r_a$ of the shock front in the pure adiabatic Sedov
solution is given by
\begin{equation}
r_a = \beta_a (E_0t^2/\rho_0)^{1/5},
\label{eq:ra}
\end{equation}
where $\beta_a \sim 1.15$, $E_0$ is the energy input and $\rho_0$ is
the background density of the medium.  For the isotropic conduction
problem (middle column of Fig.~\ref{fig:bw3} and \ref{fig:bw80}), let
us consider a diffusivity of the form $\chi = \chi_0 \rho^a T^b$. In
our case, $a=0$ and $b=5/2$. For these parameters, the position $r_c$
of the conduction front is given by \citep{Zeldovich1966,
  Barenblatt1981}
\begin{equation}
r_c = \left[ \beta_c \chi_0 \left( \frac{E_0}{c_v \rho_0} \right)^b t \right]^{1/(nb+2)},
\label{eq:rc}
\end{equation}
where $\beta_c$ is defined by
\begin{equation}
 \beta_c = \frac{2(nb+2)}{b} \left[ \frac{2}{S_n} \frac{\Gamma(\frac{n+2}{2} + \frac{1}{b})}{\Gamma(\frac{n}{2})\Gamma(1+\frac{1}{b})} \right]^b,
 \label{eq:bc}
\end{equation}
and $n$ is the dimensionality of the problem. $\Gamma$ is Euler's
gamma function and $S_n = 1$, $2\pi$, and $4\pi$ for 1,2 and 3
dimensions, respectively.  Therefore, in 3D, $r_a \propto t^{2/5}$ and
$r_c \propto t^{2/19}$, implying that initially the conduction front
outpaces the shock front. Thus, at early times, the hydrodynamic
forces, which react on longer time scales, are negligible.  Therefore,
the density behind the conduction front is constant and only decreases
by a little amount behind the front. Conduction also makes the
temperature profile shallower.

However, at late times (Fig.~\ref{fig:bw80}) the shock front overtakes
the conduction front and the position of the shock is exactly at the
same position as in the adiabatic run, which means that the solution
is now governed by hydrodynamics. The conduction effects are limited
to the central regions behind the shock front, where the temperature
is still approaching an isothermal configuration and the density does
not fall off as rapidly as in the adiabatic run. The density,
temperature and pressure profiles all compare well with similar
simulations performed by \citet{Shestakov1999}.

\begin{figure}
\begin{center}
\includegraphics[scale=0.47]{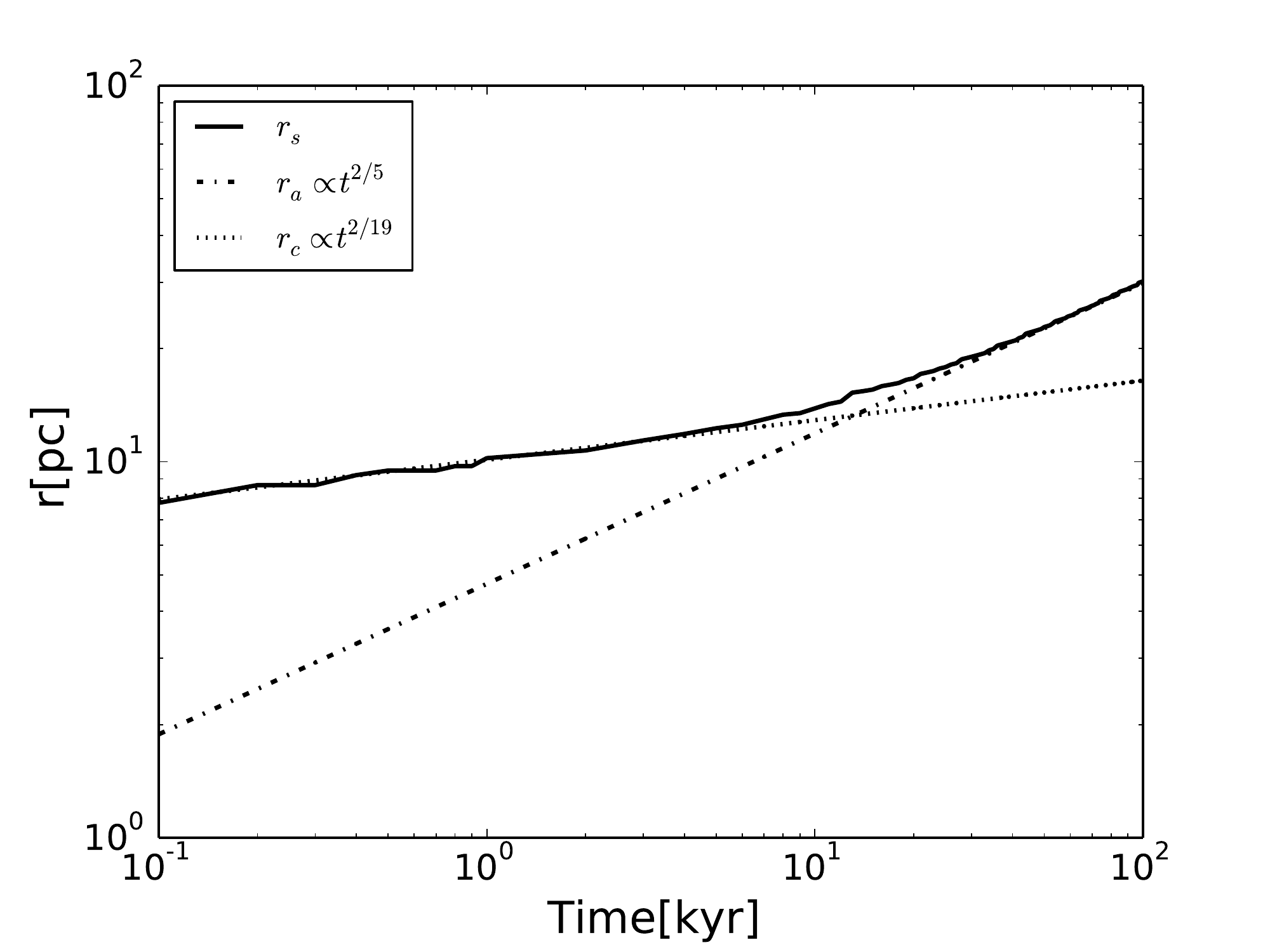}
\caption{Evolution of the conduction/shock front in the point
  explosion simulation with isotropic conduction ($r_s$ -- solid
  curve). The dotted line is the analytic expectation for a pure
  conduction problem ($r_c$), and the dot-dashed line plots the
  analytic expectation for the classic adiabatic Sedov-Taylor
  problem ($r_a$).}
\label{fig:evr}
\end{center}
\end{figure}

This dominance of different physical phenomena at different times can
be seen more clearly by plotting the radius $r_s$ of maximum density
in our isotropic conduction simulation and comparing it to the
analytic estimates for the radius $r_a$ of the shock front
(Eqn.~\ref{eq:ra}) and the radius $r_c$ of the conduction front
(Eqn.~\ref{eq:rc}) as in Fig.~\ref{fig:evr}. We can clearly see that
for $t < 4$ kyr, the density peak is almost exactly in the position of
the analytic conduction front, while for $t > 30$ kyr the density peak
is at the position of the adiabatic shock front.  During the
transitionary period both conduction and hydrodynamics play an
important role. Neither process is completely dominant, instead they
interfere constructively and advance the shock front farther than
 the analytic expectation from either process.

\subsubsection{Point explosion with anisotropic thermal conduction}

\begin{figure}
\begin{center}
\includegraphics[scale=0.47]{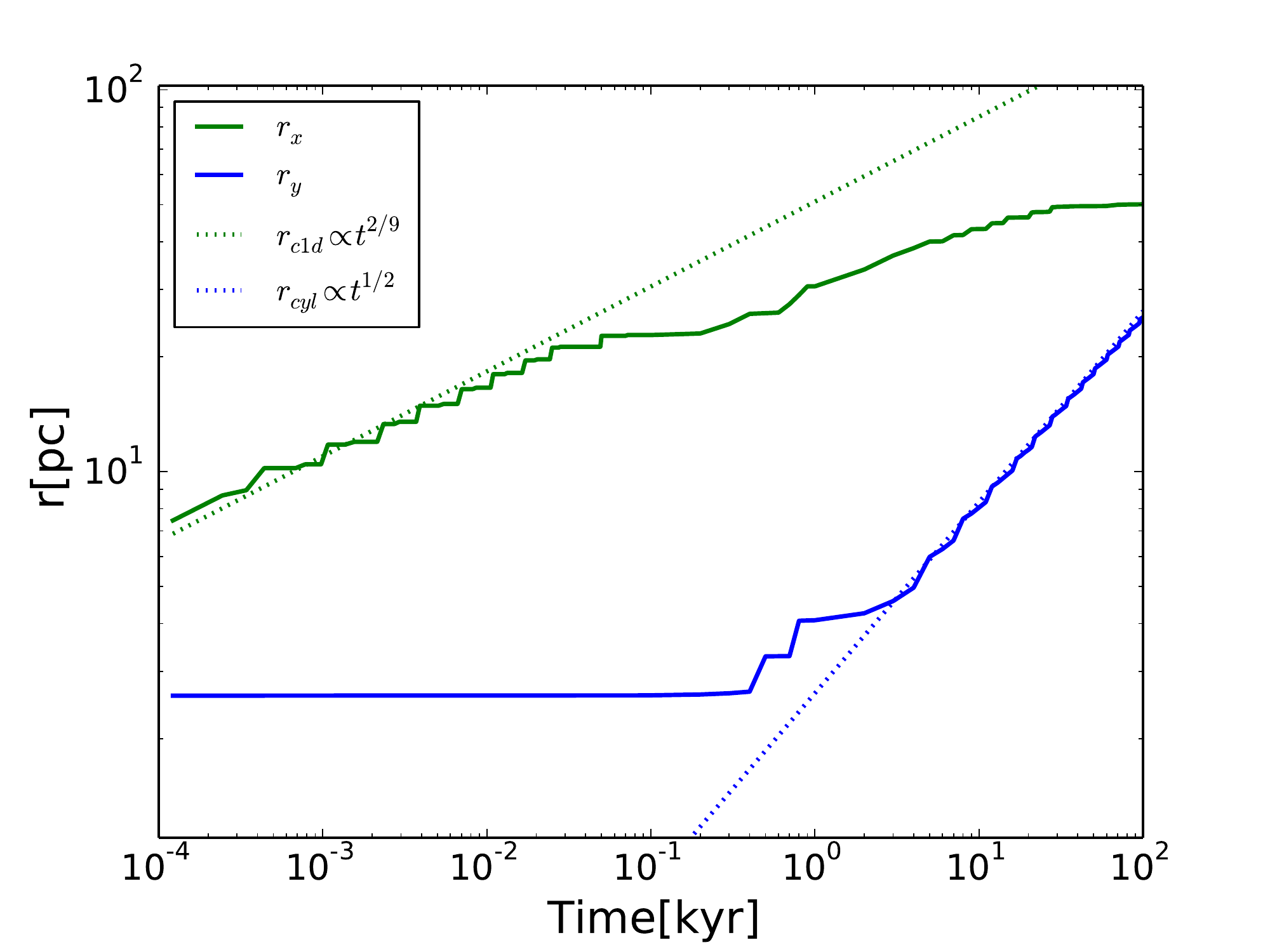}
\caption{Evolution of the conduction/shock front in the point
  explosion simulation with anisotropic conduction in the direction of
  the magnetic field ($r_x$ -- solid blue curve) and perpendicular to
  it ($r_y$ -- solid green curve).  Over-plotted is the analytic
  solution for the evolution of a pure conduction front in 1D
  ($r_{c1d}$ -- dotted blue curve) and a pure adiabatic shock front in
  2D ($r_{\rm cyl}$ -- dotted green curve).}
\label{fig:evr_aniso}
\end{center}
\end{figure}

The right columns of Fig.~\ref{fig:bw3} show the pressure, density and
temperature profiles in the simulation with anisotropic conduction at
3 kyr. The green points plot the profiles in the direction of the
magnetic field while the blue points plot the profiles perpendicular
to the magnetic field. We can see that the evolution is fundamentally
different along different axes. Along the magnetic field, the
conduction front races along very rapidly and at a much faster pace
than in the isotropic case, because the energy is diffused along a thin line rather than in a sphere.  Eqn.~\ref{eq:rc} tells us that the position of the
conduction front in 1D is $r_{c1d} \propto t^{2/9}$.  Therefore,
initially, the conduction front expands very quickly along the
magnetic field direction. Fig.~\ref{fig:evr_aniso} shows the evolution
of the conduction/shock front in the simulation along the magnetic
field axis ($r_x$, solid green curve), with the analytic expectation
overplotted (dotted green line).  We can clearly see that at very
early times the conduction front almost exactly follows the analytic
expectation for a 1D conduction problem. However, after about
$\sim 100 \,{\rm{yr}}$, the conduction front slows down because of the
advection of gas perpendicular to the magnetic field behind the
conduction front.

Perpendicular to the magnetic field the shock seems to lag behind the
classical adiabatic solution. This phenomenon was also seen in the
simulations performed by \citet{Dubois2015}. The solid blue curve of
Fig.~\ref{fig:evr_aniso} shows the evolution of the shock front
perpendicular to the magnetic field. Initially, there is very little
advection as conduction dominates most of the dynamics of the
gas. After about $\sim 1\,{\rm kyr}$, advection starts to
dominate. However, by this time the geometry of the problem has
changed from a spherical to a cylindrical blast wave. The position of
the shock front in the adiabatic cylindrical blast wave is
$r_{\rm cyl} \propto t^{1/2}$, which is denoted by the dotted blue
line in Fig.~\ref{fig:evr_aniso}. We clearly see that at late times
the evolution of the shock front in the perpendicular direction
follows the analytic expectation from a cylindrical shock. We fully
expect the radii in both cases to eventually meet at a certain time
and, afterwards the shock evolves as a regular 3D shock wave (once the
conduction timescale has become subdominant to the advection
timescale).

These results notably verify the accuracy and reliability of our
conduction scheme on randomly oriented, moving meshes. They also
validate the accuracy of the coupling between the hydrodynamics and
heat conduction in our implementation.

\subsection{Convective instabilities in a rapidly conducting  plasma}
\label{sec:intb}

The dynamics of conducting plasmas differ from that of an 
adiabatic fluid \citep{Balbus2000, Quataert2008}.  When the conduction timescale 
is much smaller than the dynamical time of the plasma (rapid conduction limit), 
the temperature gradient and the local orientation of the magnetic field 
determine the plasma's convective stability.  When the temperature increases 
with height, the convective instability is known as the heat-flux-driven buoyancy 
instability (HBI) and when it decreases with height it is described as 
magneto-thermal instability (MTI). 
 
\citet{Balbus2000} investigated weakly magnetized plasma with
${\rm d}T/{\rm d}z<0$ and identified that a small perturbation
applied to a system in hydrostatic and thermal equilibrium can induce
vigorous convection (MTI). In the rapid conduction regime, the gas is
isothermal along the magnetic field lines. Since the temperature
decreases with height, fluid elements displaced in the upward direction are warmer
than their new surroundings, making them expand and rise. Similarly,
fluid elements displaced in the downward direction
sink. \citet{McCourt2011} showed that the MTI drives sustained turbulence
for as long as the temperature gradient persists. The plasma never
becomes stable, and the magnetic field and fluid velocities
are nearly isotropic at late times.
 
\citet{Quataert2008} showed that a stable system can become unstable
even when ${\rm d}T/{\rm d}z>0$, in the presence of a background heat
flux (HBI). This can happen if the magnetic fields lines are parallel
to the temperature gradient. The regions of converging and diverging
magnetic field lines fields correspond to regions where the plasma is
locally heated and cooled. As a result, when ${\rm d}T /{\rm d}z > 0$,
a fluid element displaced downward is conductively cooled, making it
colder than the surroundings and letting it sink further down. On the
other hand, a fluid element displaced upwards becomes warmer than the
surroundings and buoyantly rises. It was shown that HBI has the tendency to re-orient vertical magnetic fields
into a horizontal configuration and can thus significantly reduce conductive heat transport in the system.
 
In this subsection, we present simulations of the MTI and HBI
instabilities using our implementation of the anisotropic thermal
conduction and compare our results to previous simulations performed
by~\citet{Parrish2005}, \citet{Parrish2008} and \citet{McCourt2011}.

\begin{figure*}
\begin{center}
\includegraphics[scale=0.41]{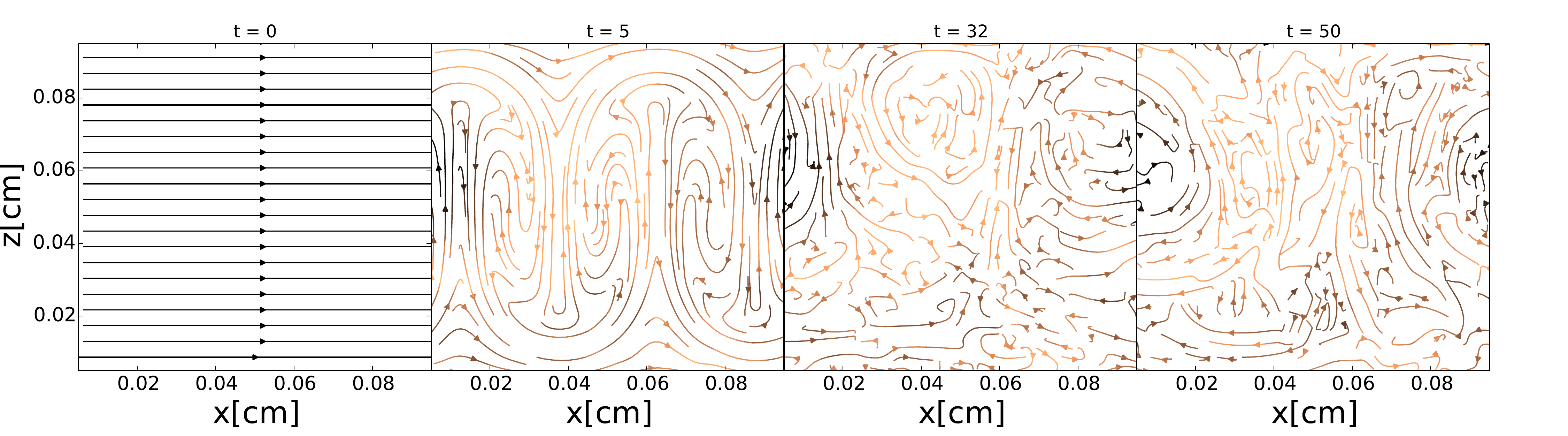}
\caption{Evolution of the magnetic field in the MTI test simulation in
  2D. We start with an initially horizontal magnetic field. The
  colours represent the magnetic field intensity at each point, with
  lighter colours depicting higher magnetic field strengths. The
  simulation time at the top of each panel is in units of the buoyancy
  timescale $t_{\rm buoy} \sim 1.7\,{\rm s}$. }
\label{fig:MTI}
\end{center}
\end{figure*}
 
\begin{figure}
\begin{center}
\includegraphics[scale=0.45]{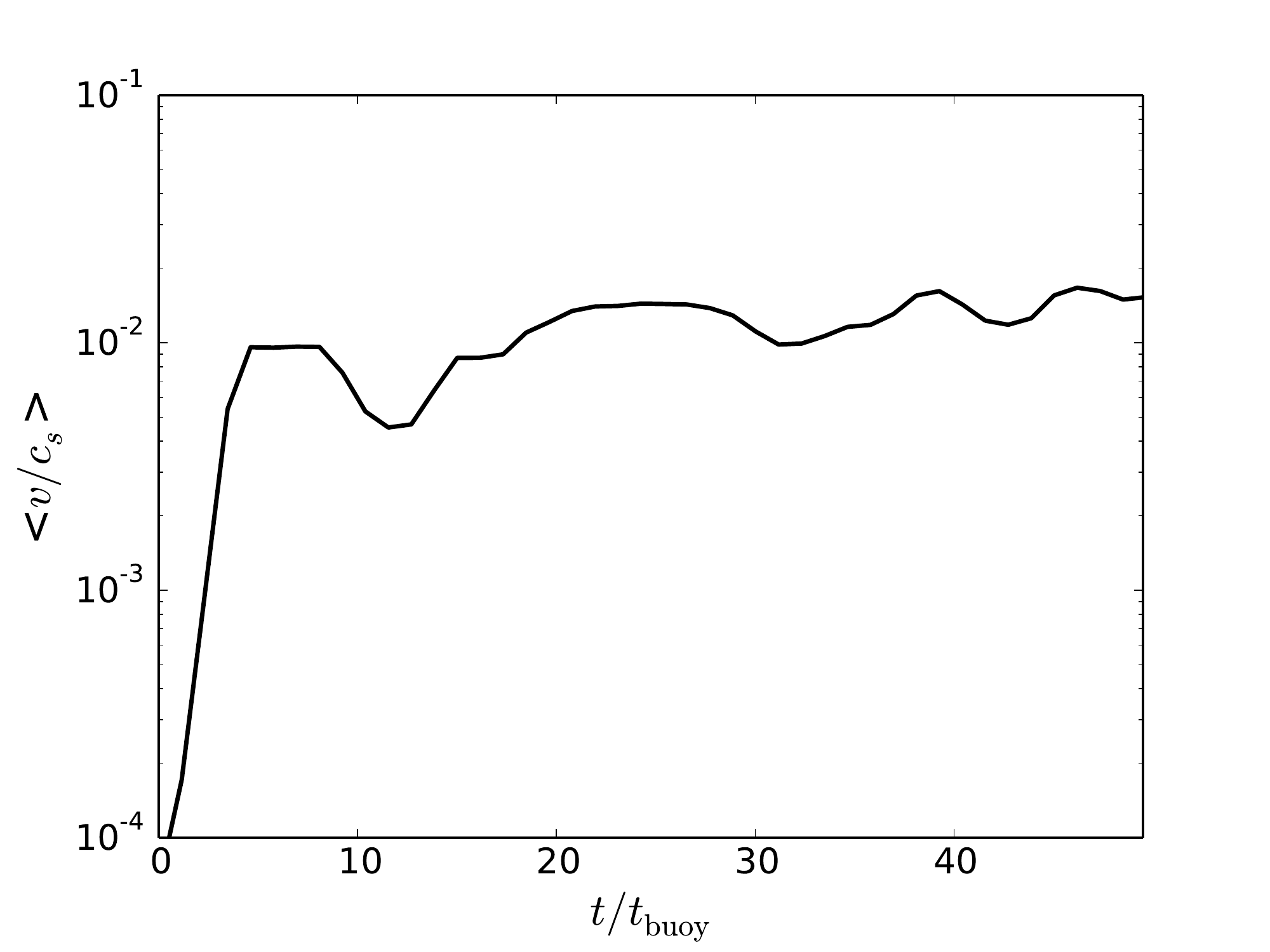}
\caption{Evolution of the volume averaged Mach number of the
  turbulence generated in the MTI simulation.}
\label{fig:MTI_mach}
\end{center}
\end{figure}
 \begin{figure}
\begin{center}
\includegraphics[scale=0.70]{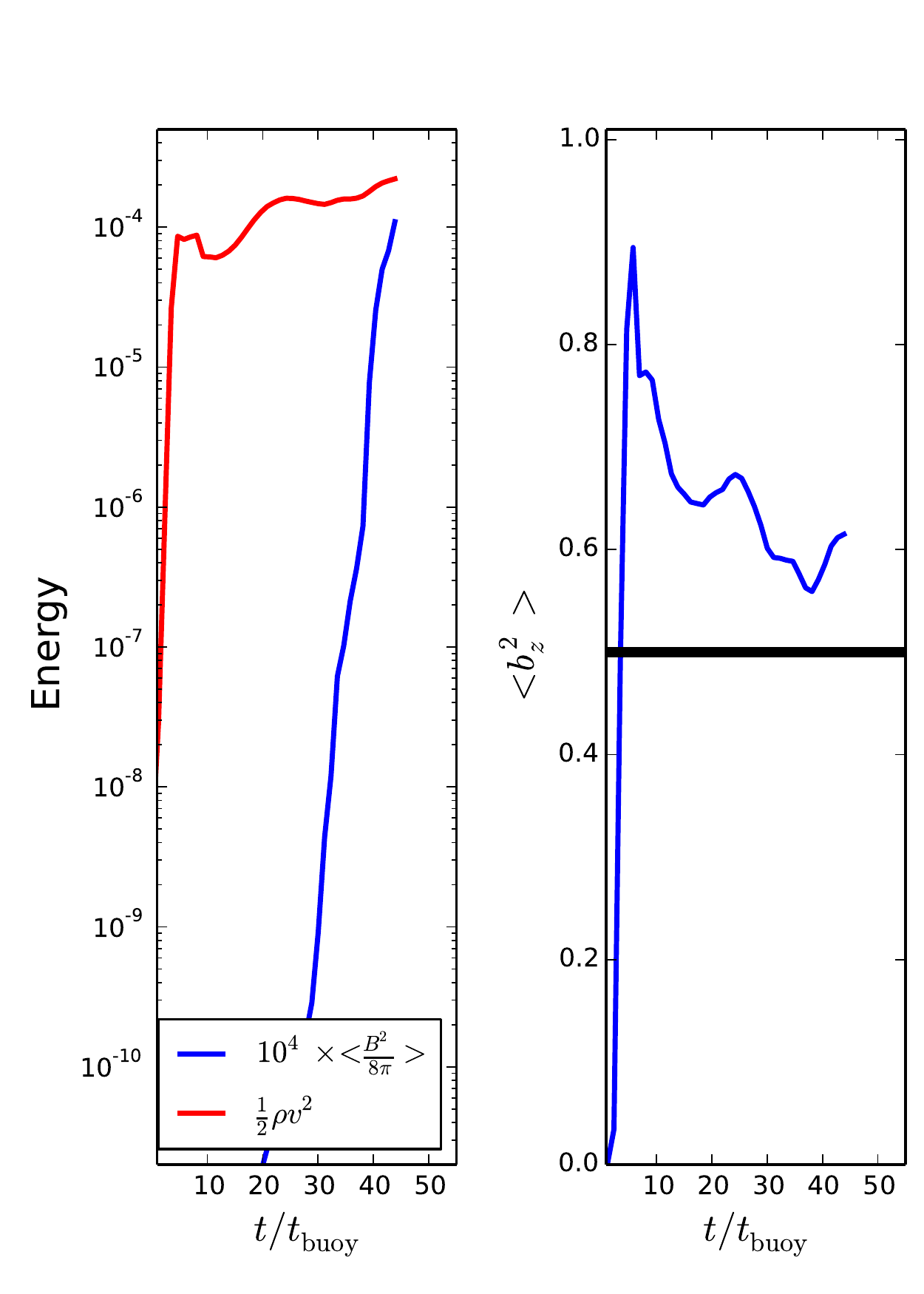}
\caption{{\em Left panel:} Evolution of the kinetic (red curve) and
  magnetic (blue curve) energy in the MTI simulation. {\em Right
    panel:} Evolution of the orientation of the magnetic field in the
  same simulation.}
\label{fig:MTI_B}
\end{center}
\end{figure}

\subsubsection{Simulation of the MTI}

We perform a 2D simulation of the MTI instability. This simulation is
local in the sense that the domain size $L$ is much smaller than the
plasma scale height $H$.  We start with a stable system in hydrostatic
and thermal equilibrium. The gravitational force field is
$\vec{g} = -g_0\vec{\hat{z}}$, with $g_0=1 \ \rm{cm\,s^{-2}}$. The internal
energy, density and the pressure profiles are set-up as in the local
simulations of \citet{Parrish2005},
\begin{equation}
\begin{split}
u &= u_0 (1-z/H), \\
\rho &= \rho_0 (1-z/H)^2, \\
P &= P_0 (1-z/H)^3, \\
\end{split}
\end{equation}
where $u_0 = 1.5 \ \rm{erg\,g^{-1}}$, $\rho_0 = 1 \ \rm{g\,cm^{-3}}$,
$P_0 = 1 \ \rm{dyne\,cm^{-2}}$ and $H=3$. We set the adiabatic index
to $\gamma = 5/3$.  The size of the computational domain is
$(0.1 \ \rm{cm})^2$, which corresponds to $L/H = 1/30$. The resolution
is $100^2$ particles. The diffusivity is set to
$\chi = 0.01 \ \rm{cm^2\,s^{-1}}$.  The initial magnetic field is set
to $\vec{B} = \{ 1, 0 \} \ \rm{nG}$.
 
\begin{figure*}
\begin{center}
\includegraphics[scale=0.4]{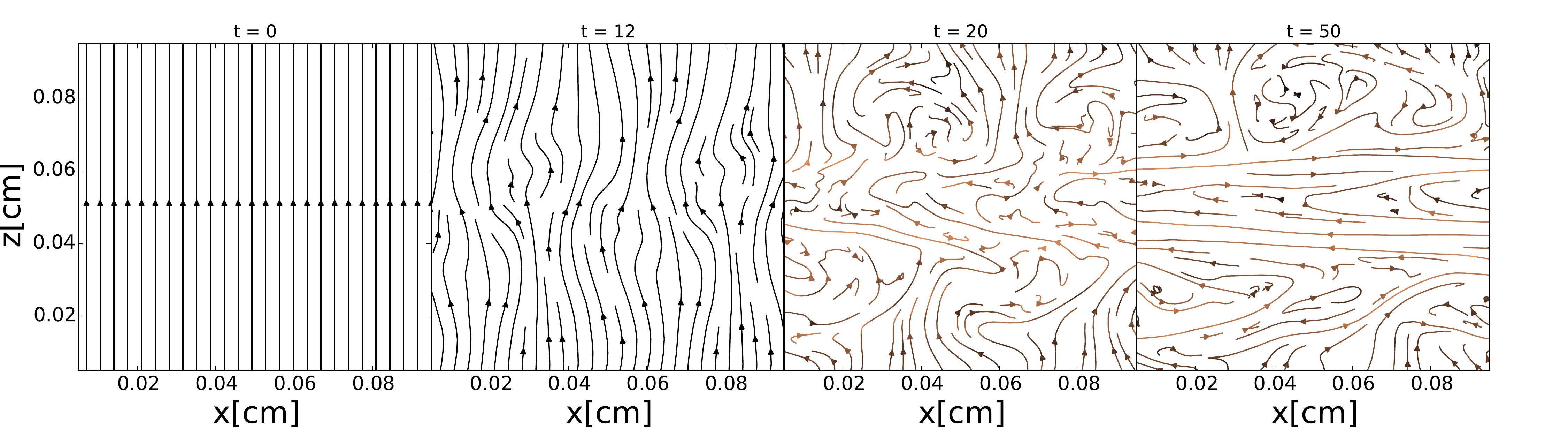}
\caption{Same as Fig.~\ref{fig:MTI}, but for the HBI simulation. The
  buoyancy time for this simulation is $t_{\rm{buoy}} \sim 1.4\,{\rm s}$.}
\label{fig:HBI}
\end{center}
\end{figure*}

We use reflective boundary conditions. The temperature at the upper and lower boundaries are constant, meaning
that they act as sources/sinks of heat during the simulation. The
density at the boundary is extrapolated from the cells below, such
that they maintain hydrostatic equilibrium throughout the duration of
the simulation. This extrapolation along with the reflecting boundary
conditions influence the evolution of the instability. To reduce this
effect we sandwich the anisotropically conducting region between
isotropic buoyantly neutral layers as in \citet{Parrish2005}.  The
domain is divided into three equal regions of length $L/3$, and the top
and bottom layers have isotropic conduction, while the middle
layer has fully anisotropic heat transport. The setup has a positive
entropy gradient meaning that the system is stable in the absence of
anisotropic conduction. All the results for the MTI are quoted for
quantities within the anisotropic conduction region. An initial
velocity perturbation of the from
\begin{equation}
v_z = 10^{-4}c_s \sin \left(\frac{4\pi x}{L}\right)
\end{equation}
is applied. We run this simulation setup with MHD and heat conduction
until $t = 50 \ t_{\rm buoy}$, where
$t_{\rm buoy} = \omega_{\rm buoy}^{-1} = |g \ \delta \ln T/\delta
z|^{-1/2} \sim 1.7\,{\rm s}$.
  
\begin{figure}
\begin{center}
\includegraphics[scale=0.45]{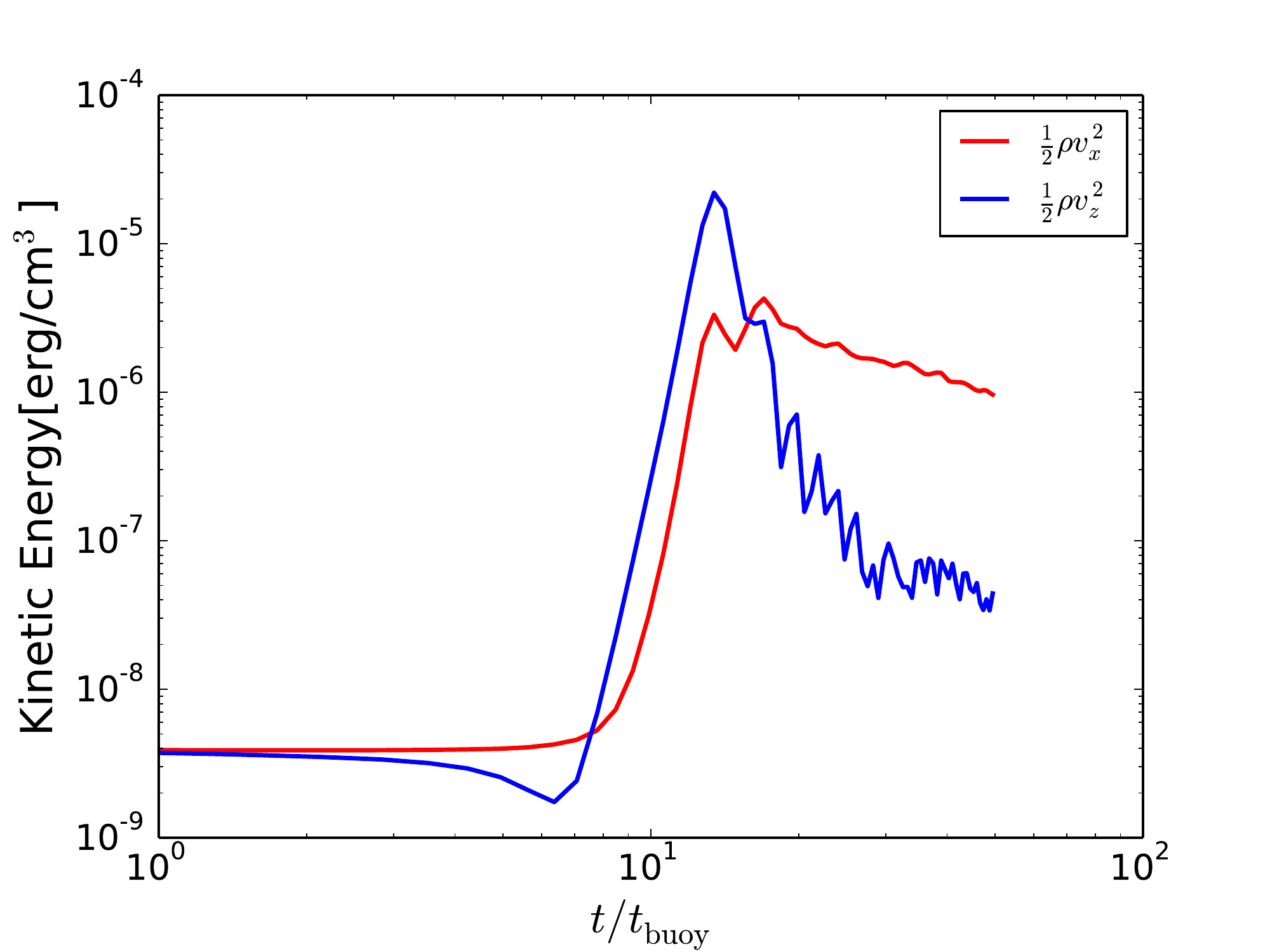}
\caption{Evolution of the vertical (blue curve) and horizontal (red
  curve) kinetic energy in the HBI simulation. The kinetic energy is
  in units of ${\rm erg\,cm^{-3}}$ and the time is in units of
  $t_{\rm buoy} \sim 1.4\,{\rm s}$.}
\label{fig:HBI_KE}
\end{center}
\end{figure}

\begin{figure}
\begin{center}
\includegraphics[scale=0.70]{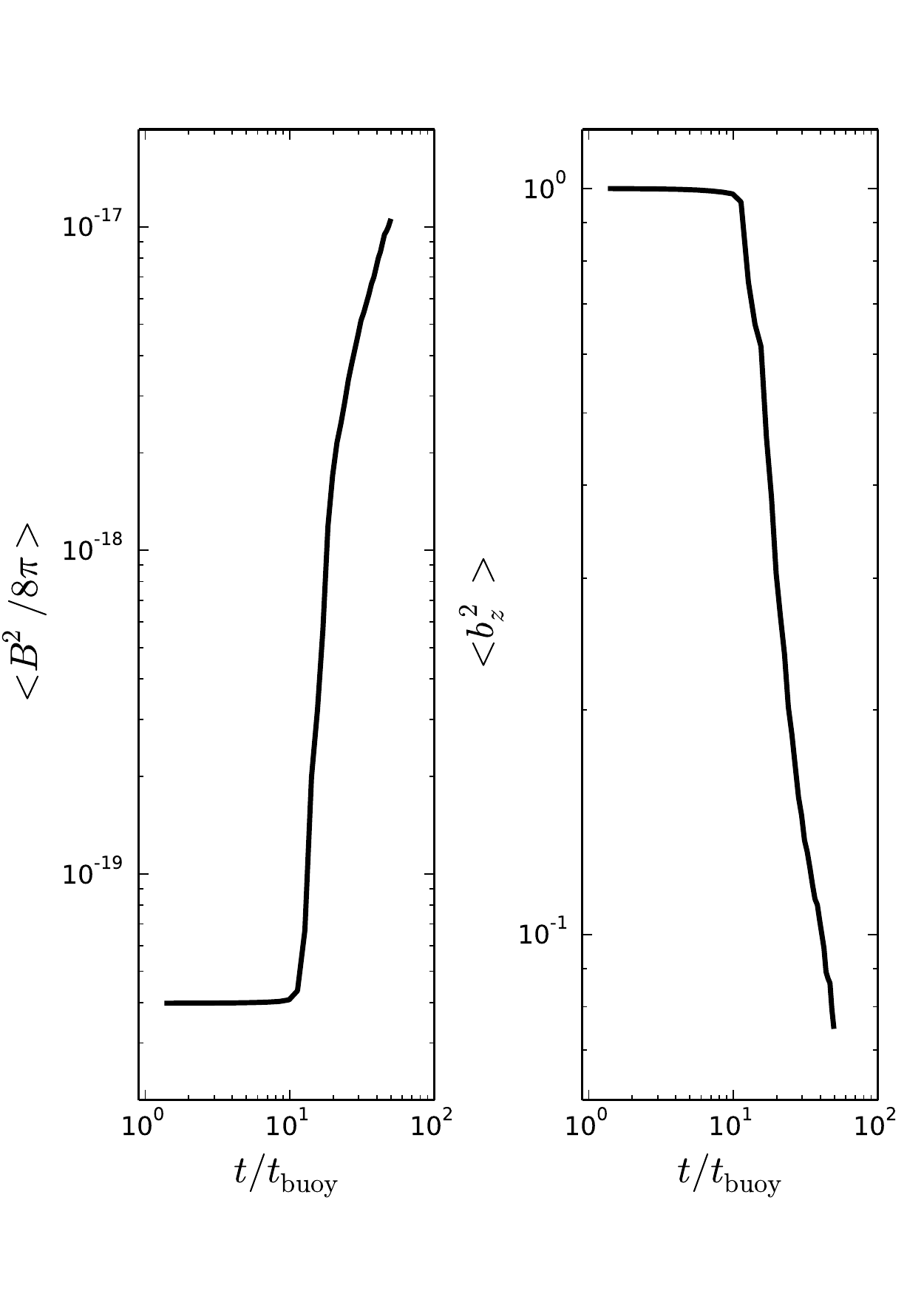}
\caption{The panel on the left hand side shows the increase of the
  magnetic energy in the HBI simulation, while the right hand side
  panel plots the evolution of the orientation of the magnetic field.}
\label{fig:HBI_B}
\end{center}
\end{figure}

Fig.~\ref{fig:MTI} shows the orientation of the magnetic field lines
as a function of time. Initially the magnetic fields are horizontal.
Within $5 \, t_{\rm buoy}$ the system becomes unstable and the
magnetic field is dragged along rising and sinking parcels of gas
making it nearly vertical.  \citet{Quataert2008} showed that the maximum growth rate of
the MTI goes to zero when $\hat{b}_z=1$. One would therefore expect
the instability to saturate at this point. This is however not the case. \citet{McCourt2011}
showed that the displacements orthogonal to
gravity generate a horizontal magnetic field component from the
vertical magnetic field. This further seeds the instability and closes the
dynamo loop. Therefore, this process continuously drives the MTI and sustains
turbulence. This is only true due to the constant temperature boundary
conditions that are imposed. If the temperatures were free to vary, the
MTI could saturate by making the plasma isothermal.

Figure~\ref{fig:MTI_mach} shows the volume-averaged Mach number within
the anisotropic region of the simulation domain as a function of
time. The increase of the average Mach number is due to MTI-driven
turbulence. The value of the average Mach number is quite similar to
the value obtained by \citet{McCourt2011} in their local 3D MTI
simulation. Of course, the Mach number is suppressed due to the small
size of the domain, increasing the domain size can potentially induce
turbulence of the order of the sound speed $v\sim c_s$.

The left hand panel of Fig.~\ref{fig:MTI_B} shows the kinetic (red
curve) and magnetic (blue curve) energies as a function of time. The
figure clearly demonstrates the non-linear amplification of the
magnetic field due to the turbulence generated by the MTI. The kinetic
energy increases rapidly at early times and then slowly saturates
after about $5 \, t_{\rm buoy}$, while the magnetic field is amplified
exponentially and will only saturate once it reaches
equipartition. The simulation needs to be run for a very long time in
order to see the saturation of the magnetic field.

The right hand panel of Fig.~\ref{fig:MTI_B} shows the evolution of
the orientation of the magnetic field as a function of time. Initially
the magnetic fields get stretched into an almost vertical
configuration. The persistent convective turbulence then drives down
the value of $\hat{b}_z$, making it asymptote to the isotropic value.
These results are quantitatively and qualitatively similar to the
results obtained in the simulations of \citet{McCourt2011}.

\subsubsection{Simulation of HBI}

Finally, we perform local 2D simulations of the HBI. The values for
the domain size, resolution, heat conduction coefficient and gravity
are the same as the ones used in the MTI simulations. We also use the
same boundary conditions. The internal energy, density and pressure
profiles are set to
\begin{equation}
\begin{split}
 u &= u_0 (1+z/H), \\
\rho &= \rho_0/(1+z/H)^3, \\
P &= P_0/(1+z/H)^2, \\
\end{split}
\end{equation}
with $u_0 = 1.5 \ \rm{erg\,g^{-1}}$, $\rho_0 = 1 \ \rm{g\,cm^{-3}}$,
$P_0 = 1 \, \rm{dyne\,cm^{-2}}$ and $H=2$. This corresponds to
$L/H = 0.05$ and $t_{\rm buoy} \sim 1.4\,{\rm s}$. The magnetic field
is initialised to $\vec{B} = \{0,1\}\,{\rm nG}$. \citet{Quataert2008}
showed that non-zero $\hat{k}_x$ and
$\hat{k}_z$, generated converging and diverging field lines that triggered HBI. Therefore the initial perturbation is set to
\begin{equation}
\vec{v} = 10^{-4}c_s \left[\sin\left(\frac{3\pi y}{L}\right) \hat{i} + \sin\left(\frac{4\pi x}{L}\right) \hat{k}\right].
\end{equation}

As in the previous subsection, all the results for the HBI are quoted
for quantities within the anisotropic conduction region.
Fig.~\ref{fig:HBI} shows the orientation of the magnetic fields as a
function of time. By
$t = 10\, t_{\rm buoy}$, the magnetic field lines get perturbed enough
that there are regions where they converge or diverge. These
correspond to regions where the plasma is locally heated and
cooled. As a result, when ${\rm d}T /{\rm d}z > 0$, a fluid element
displaced downward is conductively cooled via the background heat
flux, making it colder than the surroundings and therefore letting it
sink further down. On the other hand, a fluid element with an upward
displacement gains energy, becomes warmer than the surroundings and
buoyantly rises. The evolution becomes non-linear by
$t \sim 12 \, t_{\rm buoy}$.  Afterwards the
instability saturates and the magnetic field settles into an almost
horizontal configuration by $t = 50\, t_{\rm buoy}$.
 
This behaviour can be more clearly seen in Fig.~\ref{fig:HBI_KE},
which shows the evolution of the vertical (blue curve) and horizontal
(red curve) kinetic energy components as a function of time. During
the linear growth phase of the instability, both components of the
velocity field accelerate to equipartition with each other, which
takes place at about $t \sim 12\, t_{\rm buoy}$.  As the instability
saturates the plasma becomes buoyantly stable and suppresses vertical
motions whereas the horizontal motions continue unperturbed. 
 
The left panel of Fig.~\ref{fig:HBI_B} shows the evolution of the mean
magnetic field energy as a function of time. The linear growth ends at
$t \sim 12 \, t_{\rm buoy}$, and most of the evolution of the magnetic
field happens afterwards. This evolution is driven by the horizontal motions which both amplify and reorient the
magnetic field. After a brief period of exponential growth, the field
amplification is roughly linear in time. The right panel of
Fig.~\ref{fig:HBI_B} shows the evolution of the magnetic field
orientation as a function of time. As expected, the HBI converts
vertical magnetic fields into horizontal ones.  Quantitatively, we
expect that $\hat{b}_z \propto t^{-1}$, which is quite close to the
value $\hat{b}_z \propto t^{-0.86}$ obtained in our simulation.  The 
reorienting of the magnetic field also has the effect of severely reducing
the conductive heat flux through the plasma.  
Stretching the field lines in this manner amplifies the field, and in
our simulations the magnetic field strength is given by
$B^2 \propto t^{1.83}$, which is again close to the expected value of
$B^2 \propto t^2$. The saturated state of the HBI is buoyantly stable. We reiterate
previous results and state that the HBI has the tendency to insulate
conductive and convective transport in plasmas, which may
significantly reduce the amount of conductive heat transport in
clusters of galaxies.

\section{Conclusions}

In this work, we have presented an implementation of an extremum
preserving anisotropic diffusion solver that works well on the
unstructured moving mesh of {\sc Arepo}.  The method relies on
splitting the one-sided facet fluxes into normal and oblique
components and on limiting the oblique fluxes such that the method is
locally conservative and also extremum preserving. The scheme makes
use of harmonic averaging points and proposes a very simple yet robust
interpolation scheme that works well for strong heterogeneous and
anisotropic diffusion problems. In addition, the required
discretisation stencil is essentially small.  We also present
efficient fully implicit and semi-implicit time integration schemes in
order to overcome the restrictive nature of the diffusion time step,
which arise due to the $\Delta x^2$ dependance of the von Neumann
stability condition.

We have tested this implementation on a variety of numerical
problems. The method works really well overall and reproduces analytic
results very well in the test of the diffusion of a temperature step
function. The main drawback of the method seems to be that the
convergence order for highly anisotropic problems like the diffusion
of a hot patch of gas in a circular magnetic field is only about
$\sim 0.4-0.5$. In future work, we will try to improve upon this by
choosing higher order interpolation schemes. Coming up with a less
restrictive flux limiter should also help. We also verify that the
numerical perpendicular diffusion is about $10^{-3}$ for
$\Delta x/L \sim 0.01$, comparable to the values obtained by other
methods \citep{Sharma2007}. We also obtain a good
numerical convergence rate for the perpendicular diffusion, of the
order of $1.5$.
 
We have also verified the accuracy and robustness of the code when
conduction and hydrodynamics are coupled by simulating point
explosions with isotropic and anisotropic heat conduction. We show
that the point explosion test with isotropic heat conduction can be
split into two regimes, the conduction dominated regime and the
advection dominated regime. Conduction dominates at early times and
accelerates the shock front beyond the classical Sedov solution.  At
late times, the hydrodynamic forces take over and the position of the
shock front returns to that of the Sedov solution. The temperature
profile behind the shock front is much shallower than the adiabatic
case. This has consequences, for example, for observations attempting
to diagnose hot winds with X-ray radiation. In particular,
\citet{Strickland1997} find a flat temperature gradient potentially
consistent with our simulations in the M82 superwind on kpc scales.

We also a perform Sedov blast wave explosion test with anisotropic
thermal conduction. Initially, the conduction front races along the
magnetic field while there is very little advection perpendicular to
it. The position of the conduction front along the magnetic field
accurately follows the solution of a 1D conduction problem. At late
times, the advection perpendicular to the magnetic field decelerates
the conduction front. This also means that the geometry of the problem
changes from a spherical to a cylindrical shock. When
advection does start to dominate, the simulation accurately follows the
cylindrical Sedov solution. We conclude that the ISM magnetic fields
play an important role in the evolution of SN remnants.

Finally, we simulate convective instabilities induced by anisotropic
conduction in a rapidly conducting plasma. We simulate the
magneto-thermal and heat-flux-driven buoyancy instability in 2D and
verify the results of \citet{Parrish2005, Parrish2008,
  McCourt2011}. We find that, in the MTI, the initially horizontal
magnetic field is quickly disrupted, the
system becomes unstable and the magnetic field is dragged along rising
and sinking parcels of gas making the field orientation nearly
vertical.  Horizontal displacements makes the system non-linearly
unstable to the MTI thereby generating horizontal magnetic fields and
so on. At late times the magnetic field is completely randomised. This
turbulence also increases the magnetic and the kinetic energy of the
gas and thus generates additional pressure support against
gravity. This pressure support might become important, especially in
cluster environments \citep{McCourt2013}. We also simulate HBI
instability and get similar B field magnification and KE increase
values as previous simulations. However, the HBI quietly saturates
when the magnetic field becomes horizontal. Therefore, HBI tends to
hinder conductive transport in clusters.

To summarise, we have implemented an efficient, robust, extremum
preserving anisotropic thermal conduction solver in {\sc Arepo}. This
algorithm can be easily generalized to simulate other interesting
astrophysical diffusion processes as well, such as cosmic ray
diffusion \citep{Pfrommer2013} or radiative transfer problems. In the
latter case, it can be used for moment-based radiative transfer
calculations (such as in \citealt{Gnedin2001, Petkova2009,
  Rosdahl2015}) where the anisotropic diffusion tensor is replaced by
a local Eddington tensor, encoding information about the preferred
local propagation direction of the photons from a set of discrete
sources.

In forthcoming work, we plan to use the implementation introduced here
to study timely problems in astrophysics related to conduction, such
as heat transport in clusters, the evolution of supernova remnants,
role of convective instabilities in galaxy clusters, and the
distribution of heat from AGN jets into the ICM. An extension to
cosmic ray transport is another obvious research direction where the
anisotropic diffusion solver could be fruitfully applied.

\section*{Acknowledgements}
The simulations were performed on the joint MIT-Harvard computing
cluster supported by MKI and FAS. VS and RP acknowledge support by the
European Research Council under ERC-StG grant EXAGAL 308037 and by the
Klaus Tschira Foundation. MV acknowledges support through an MIT RSC
award.  \bibliographystyle{mnras} \bibliography{ms.bib}

\label{lastpage}

\end{document}